\documentclass[11pt,letterpaper]{article}
\pdfoutput=1
\usepackage{jheppub}
\usepackage{bbm}
\usepackage{mathrsfs}
\usepackage{slashed}
\usepackage{caption}
\usepackage{epstopdf}
\usepackage[normalem]{ulem}
\usepackage[bottom]{footmisc}
\usepackage{subcaption}
\usepackage{bbold}
\usepackage{titlesec}
\usepackage{threeparttable}
\usepackage{booktabs}
\usepackage{changepage}
\usepackage[utf8]{inputenc}
\usepackage{dsfont}
\usepackage{changes}
\usepackage{grffile}
 
\usepackage{graphicx}  
\usepackage{dcolumn}   
\usepackage{bm}        
\usepackage{amssymb}   
\usepackage{setspace}
\usepackage{amsmath, amssymb, setspace}
\usepackage{array}
\usepackage{booktabs}
\usepackage{caption}
\usepackage{indentfirst}
\usepackage{float}
\usepackage{lmodern}
\usepackage{multirow}
\usepackage{soul}
\usepackage[normalem]{ulem}
\usepackage{xcolor}

\newcommand{\aln}[1]{\begin{align}#1\end{align}}

\allowdisplaybreaks

\title{
Q-Balls in the presence of attractive force
}

\author[a,b]{Yu Hamada,}
\author[c]{Kiyoharu Kawana,}
\author[c]{TaeHun Kim,}
\author[c,d]{Philip Lu}

\affiliation[a]{Deutsches Elektronen-Synchrotron DESY, Notkestr. 85, 22607 Hamburg, Germany}
\affiliation[b]{Research and Education Center for Natural Sciences, Keio University, 4-1-1 Hiyoshi, Yokohama, Kanagawa 223-8521, Japan}
\affiliation[c]{School of Physics, Korea Institute for Advanced Study, 85 Hoegi-ro, Dongdaemun-gu, Seoul 02455, Republic of Korea}
\affiliation[d]{Center for Theoretical Physics, Department of Physics and Astronomy, Seoul National
University, 1 Gwanak-ro, Gwanak-gu, Seoul 08826, Republic of Korea}

\emailAdd{yu.hamada@desy.de}
\emailAdd{kkiyoharu@kias.re.kr}
\emailAdd{gimthcha@kias.re.kr}
\emailAdd{philiplu11@gmail.com}

\abstract{
Q-balls are non-topological solitons in field theories whose stability is typically guaranteed by the existence of a global conserved charge. 
A classic realization is the Friedberg-Lee-Sirlin (FLS) Q-ball in a two-scalar system where a real scalar $\chi$ triggers symmetry breaking and confines a complex scalar $\Phi$ with a global $U(1)$ symmetry.
A quartic interaction $\kappa \chi^2|\Phi|^2$ with $\kappa>0$ is usually considered to produce a nontrivial Q-ball configuration, and this repulsive force contributes to its stability. 
On the other hand, the attractive cubic interaction $\Lambda \chi |\Phi|^2$ is generally allowed in a renormalizable theory and could induce an instability.
In this paper, we study the behavior of the Q-ball under the influence of this attractive force which has been overlooked. 
We find approximate Q-ball solutions in the limit of weak and moderate force couplings using the thin-wall and thick-wall approximations respectively.  
Our analytical results are consistent with numerical simulations and predict the parameter dependencies of the maximum charge.
A crucial difference with the ordinary FLS Q-ball is the existence of the maximum charge beyond which the Q-ball solution is classically unstable.
Such a limitation of the charge fundamentally affects Q-ball formation in the early Universe and could plausibly lead to the formation of primordial black holes. 
}

\begin{document}
\preprint{DESY-24-102}
\maketitle
\flushbottom

\section{Introduction}  \label{sec:intro}

Non-trivial solutions of classical equations of motion in field theories play an important role in our understanding of the non-perturbative nature of many-body systems. 
Stationary solutions called solitons have many applications and are thus widely studied in disparate fields such as particle physics, cosmology, condensed matter and so on~\cite{Manton:2004tk,Tong:2005un,Johnson_2002,Nugaev:2019vru}.

Solitons can be further categorized into topological solitons whose stability is guaranteed due to a topological property of the field configuration, and non-topological solitons whose stability depends on other mechanisms such as a conserved charge associated with a symmetry. 
A famous example in this latter category is Coleman's Q-ball~\cite{Coleman:1985ki} where a single complex scalar field can develop a nontrivial classical configuration with a finite charge by itself. In this paper, we will focus on another type of non-topological soliton based off of the Friedberg-Lee-Sirlin (FLS) Q-ball~\cite{Friedberg:1976me,Heeck:2023idx,Kim:2024vam}, in which there is an additional real scalar $\chi$ as well as a complex scalar $\Phi$ with a global $U(1)$ symmetry.
The potential of $\chi$ triggers symmetry breaking, which allows for a soliton solution with a finite charge in the false vacuum. 
In this model, a repulsive-type interaction such as $\kappa \chi^2|\Phi|^2~(\kappa>0)$ is often considered, which also plays an important role in stabilizing the Q-ball because it gives a mass for $\Phi$ in the true vacuum. 
In this case, the stability of Q-ball is simply determined by its energy profile $E(Q)$ and the mass of $\Phi$ (see section~\ref{sec:Q ball} for the details), and the repulsive interaction causes no stability issues. 
However, more general renormalizable Lagrangians also allow interactions such as  $\Lambda \chi |\Phi|^2$, which is the scalar counterpart of the Yukawa interaction and plays an attractive force among  $\Phi$ particles.
The attractive nature of the interaction crucially changes the behavior of Q-balls and can cause an instability. 
In this paper, we study the FLS Q-ball in the presence of the attractive force and clarify the essential differences compared to the traditional case. 

Q-balls can have wide-ranging consequences for cosmology, as a dark matter candidate~\cite{Kusenko:1997si,Krylov:2013qe,Huang:2017kzu,Jiang:2024zrb}, generating the baryon asymmetry of the Universe~\cite{Enqvist:1997si,Kasuya:2014ofa,Kasuya:2012mh}, and present viable pathways to primordial black hole formation~\cite{Cotner:2017tir,Cotner:2019ykd,Amendola:2017xhl,Savastano:2019zpr,Flores:2020drq,Domenech:2021uyx,Domenech:2023afs}. 
It is therefore interesting to consider the stability of FLS Q-balls with an attractive interaction. 
As we will see below, the addition of attractive force constrains the allowed value of $Q$\footnote{
Such a limitation of the charge also applies to other attractive systems such as boson stars~\cite{Shnir:2022lba, Ruffini:1969qy}, where gravitational interaction plays the role of the attractive force.}, which can have implications for cosmology regarding its abundance as dark matter and its formation during a first-order phase transition in the early Universe.

The organization of this paper is as follows:
in section~\ref{sec:qball}, we present analytical analyses of the Q-ball solutions, deriving scaling relations and the maximum stable charge, as well as discussing the stability conditions. 
In section~\ref{sec:numerical}, we detail our computational methods and present numerical results. 
Finally, we summarize our results with a discussion of possible implications for cosmology and comment on the possibility of forming PBHs by the collapse of Q-balls in section~\ref{sec:conclusions}.
Throughout the paper, we use the mostly negative signature for the metric $\eta_{\mu\nu}^{}=(+,-,-,-)$.

\section{Q-Ball}\label{sec:Q ball}
We discuss Q-ball solution in the two-scalar system in the presence of an attractive force i.e. cubic interaction. 
To understand the qualitative behaviors of Q-ball, we find analytical solutions in the thin- and thick-wall approximations.
Through this analysis, we find that the existence of the attractive interaction significantly changes the Q-ball behaviors as a function of internal phase angular frequency $\omega$ (see eq.~(\ref{eq:Qballansatz}) for the definition). 
In particular, there exists a maximum value $Q_{\rm max}^{}$ of the Q-ball charge for a given strength (coupling) of attractive force, which implies that Q-ball with $Q>Q_{\rm max}^{}$ cannot be created when the system undergoes a (first-order) phase transition in the early Universe.   

\label{sec:qball}
\subsection{Lagrangian and variational principle} \label{ssec:Lagrangian}
We introduce a Lagrangian with a real scalar $\chi$ and a complex scalar $\Phi$ 
\begin{align}
\label{eq:lagrangian}
 \mathcal{L}= \frac{1}{2}(\partial_\mu \chi)^2 + |\partial_\mu \Phi|^2 -V(\chi,\Phi)~,
\end{align}
where the potential is given by 
\begin{align}
 V(\chi,\Phi)& = U(\chi) + \Lambda \chi |\Phi|^2 + \kappa \chi^2 |\Phi|^2~. 
\end{align}
When $\Lambda=0$, this Lagrangian is the same as the classic FLS Q-ball~\cite{Friedberg:1976me}. 
The complex scalar $\Phi$ has a global $U(1)$ symmetry $\Phi\rightarrow e^{i\theta}\Phi$, which results in the conservation of the particle number 
\aln{
Q = i \int d^3 x (\Phi^\dagger \partial_0^{} \Phi - \partial_0^{}\Phi^\dagger \Phi)~. 
\label{eq:Q-def}
} 
As for the $\chi$ potential, we consider the following typical one: 
\begin{align}
 U(\chi) &= \frac{\lambda}{4!} (\chi^2 - v^2)^2~.
\end{align}
When $\Lambda\sim 0$ and $\kappa\gtrsim 0$, the true minimum exists at $(\chi,\Phi)=(v,0)$ and the masses of scalar fields are 
\aln{
m_\Phi^2 = \Lambda v + \kappa v^2~,\quad m_\chi^2 = \frac{\lambda}{3} v^2~.
}
The coupling $\Lambda$ explicitly breaks the $Z_2$ symmetry of $\chi \rightarrow - \chi$, shifting the potential minima for $\chi$ inside the Q-ball where $|\Phi| \neq 0$ while giving a correction to the mass of $\Phi$ in the true vacuum. 
Since the cubic term corresponds to an attractive interaction among $\Phi$ particles, this can modify the form of the Q-ball solution and destabilize it. 
Our objective is to find the available range of stable Q-ball configurations by solving the classical field equations for $\chi$ and $\Phi$, both analytically and numerically. 
In the following discussion, we focus on the parameter space
\aln{m_\Phi^2 = \Lambda v+\kappa v^2>0
} 
so that $\Phi$ does not develop a nonzero vacuum expectation value (VEV). 
In particular, we fix $\kappa>0$ throughout the paper.

We want a classical solution that minimizes the total energy 
\aln{
E=\int d^3x\left(\frac{1}{2}\dot{\chi}^2+\frac{1}{2}(\nabla \chi)^2+|\dot{\Phi}|^2+|\nabla \Phi|^2+U(\chi)
+\Lambda \chi |\Phi|^2+\kappa\chi^2|\Phi|^2\right)
\label{FLS Hamiltonian}
}
with a fixed charge $Q$. 
This can be done by introducing a Lagrange multiplier $\omega$ and considering the functional 
\aln{
{\cal {E}}&=E-\omega Q~
\\
&=\int d^3x\left[(\dot{\Phi}+i\omega \Phi)^*(\dot{\Phi}+i\omega \Phi)+\frac{1}{2}\dot{\chi}^2 +|\nabla \Phi|^2+\frac{1}{2}(\nabla \chi)^2 \right. \nonumber \\
&\left. \qquad \qquad \ \vphantom{(\dot{\Phi}+i\omega \Phi)^*(\dot{\Phi}+i\omega \Phi)+\frac{1}{2}\dot{\chi}+|\nabla \Phi|^2+\frac{1}{2}(\nabla \chi)^2} 
-\omega^2|\Phi|^2+U(\chi)+\Lambda \chi |\Phi|^2+\kappa \chi^2|\Phi|^2
\right]~,
\label{modified energy}
}
which shows that the first term is minimized for a stationary ansatz $\Phi=\phi(\vec{x})e^{-i\omega t}/\sqrt{2}$.  
Although a complicated angular dependence is possible for the spatial part in general, we assume a spherically symmetric form for the least-energy state,
\begin{equation}
 \chi = \chi(r), \quad \Phi = \frac{\phi(r)}{\sqrt{2}} e^{- i\omega t}~,\quad \chi(r),\phi(r)\in \mathbb{R}~.
 \label{eq:Qballansatz}
\end{equation}
With this form, the conserved charge (\ref{eq:Q-def}) becomes 
\begin{equation}
Q = 4\pi \omega \int^\infty_0 r^2 dr \, \phi^2(r)~. 
\label{eq:Q}
\end{equation}
Substituting eq.~\eqref{eq:Qballansatz} into the equation of motions (EOMs) for $\chi$ and $\Phi$ derived from the Lagrangian eq.~\eqref{eq:lagrangian}, we obtain the Q-ball EOM:
\begin{subequations}
\begin{eqnarray}
    &&\frac{1}{r^2} \frac{d}{dr} \left(r^2 \frac{d\chi}{dr} \right) - \frac{dU(\chi)}{d\chi} - \frac{\Lambda}{2} \phi^2 - \kappa \phi^2 \chi = 0~, \\
    &&\frac{1}{r^2} \frac{d}{dr} \left(r^2 \frac{d\phi}{dr} \right) + \omega^2 \phi - (\Lambda \chi + \kappa \chi^2) \phi = 0~.
\end{eqnarray}%
\label{eq:QballEOM}%
\end{subequations}
The solution of these equations is at least a stationary point of the energy functional for a given $Q$. 
We can restrict the parameter space by observing the following symmetries. From eqs.~\eqref{eq:Q} and~\eqref{eq:QballEOM}, if $(\omega, \chi(r), \phi(r))$ is a solution of eqs.~\eqref{eq:QballEOM} for a given $\Lambda$ with charge $Q$, then $(-\omega, \chi(r), \phi(r))$ is also a solution for the same $\Lambda$ and charge $-Q$. Similarly, $(\omega, -\chi(r), \phi(r))$ is also a solution for $-\Lambda$ with the same charge $Q$. The former allows us to focus only on positively charged Q-balls with $\omega > 0$, and the latter allows us to fix $\chi(\infty) = +v$, allowing $\Lambda$ to take both signs. 

A few comments are necessary. 
For any $\Lambda$, there is the so-called plane-wave solution of eqs.~(\ref{eq:QballEOM}) such that free $\Phi$ particles homogeneously exist in the true vacuum.
In this case, we have 
\aln{\omega\rightarrow m_\Phi^{}~,\quad E\rightarrow m_\Phi^{}Q
}
in the thermodynamic limit $V_3^{}\rightarrow \infty$ with $V_3$ the three-dimensional volume because each $\Phi$ particles simply oscillates with the frequency $\omega=m_\Phi^{}$.       
On the other hand, the Q-ball solution corresponds to another branch with $\omega < m_\Phi$ such that the binding of $\Phi$ particles is energetically favored compared to the free case. 
The amount of the localized charge and the size of the Q-ball are determined by the balance between this energy difference of $\Phi$ particles and the false vacuum energy of $\chi$. 
In sections~\ref{ssec:thin-wall} and \ref{ssec:thickwall}, we show analytical solutions for small and moderate values of $\Lambda$, which respectively hold in the thin and thick-wall approximations.\\

\noindent{\bf Dimensionless units.} The parameter scaling of our solutions can be made more explicit through the use of dimensionless units.
We can reduce the four parameters of the Lagrangian~\eqref{eq:lagrangian}, $\Lambda$, $\kappa$, $\lambda$, and $v$, two scalar fields $\chi$ and $\Phi$, and position $x$ to five quantities:
dimensionless coordinates $x^\mu = \sqrt{\kappa} v x^\mu$, dimensionless fields $\tilde{\chi} = \chi / v$, $\tilde{\Phi} = \Phi / v$, and dimensionless parameters $\tilde{\Lambda} = \Lambda / \kappa v$, $\tilde{\lambda} = \lambda / \kappa$. With these substitutions, the action becomes 
\begin{equation}
    S = \frac{1}{\kappa} \int d^4 \tilde{x} \left[ \frac{1}{2} (\tilde{\partial}_\mu \tilde{\chi})^2 + |\tilde{\partial}_\mu \tilde{\Phi}|^2 - \frac{\tilde{\lambda}}{24} ( \tilde{\chi}^2 - 1 )^2 - \tilde{\Lambda}\tilde{\chi}|\tilde{\Phi}|^2 - \tilde{\chi}^2 |\tilde{\Phi}|^2 \right],
\end{equation}
so we need to consider only $\tilde{\lambda}$ and $\tilde{\Lambda}$ as independent parameters in numerically solving the Q-ball EOM for dimensionless fields, 
\begin{subequations}
\begin{eqnarray}
    &&\frac{1}{\tilde{r}^2} \frac{d}{d\tilde{r}} \left(\tilde{r}^2 \frac{d\tilde{\chi}}{d\tilde{r}} \right) - \frac{\tilde{\lambda}}{6}\tilde{\chi}(\tilde{\chi}^2-1) - \frac{\tilde{\Lambda}}{2} \tilde{\phi}^2 - \tilde{\phi}^2 \tilde{\chi} = 0~, \\
    &&\frac{1}{\tilde{r}^2} \frac{d}{d\tilde{r}} \left(\tilde{r}^2 \frac{d\tilde{\phi}}{d\tilde{r}} \right) + \tilde{\omega}^2 \tilde{\phi} - (\tilde{\Lambda} \tilde{\chi} + \tilde{\chi}^2) \tilde{\phi} = 0
\end{eqnarray}%
\label{eq:QballEOMdimless}%
\end{subequations}
where $\tilde{\omega} = \omega / \sqrt{\kappa}v$. The obtained solution can be readily converted to physical fields and variables once we select $\kappa$ and $v$. This dimensionless normalization is also implemented in ref.~\cite{Heeck:2023idx}, where the cubic interaction is not considered.

The dimensionless masses of the dimensionless fields are $\tilde{m}_\chi^2 = \tilde{\lambda}/3 = m_\chi^2 / \kappa v^2$ and $\tilde{m}_\Phi^2 = \tilde{\Lambda} + 1 = m_\Phi^2 / \kappa v^2$. The charge is
\begin{equation}
    \tilde{Q} = 4\pi \tilde{\omega} \int^\infty_0 \tilde{r}^2 d\tilde{r} \tilde{\phi}^2 (\tilde{r}) = \kappa 4\pi \omega \int^\infty_0 r^2 dr \phi^2(r) = \kappa Q
\end{equation}
and the energy is
\begin{equation}
    \tilde{E} = \int d^3 \tilde{x} \left[\frac{1}{2} (\partial_{\tilde{t}} \tilde{\chi})^2 + \frac{1}{2} |\tilde{\nabla} \tilde{\chi}|^2 + |\partial_{\tilde{t}} \tilde{\Phi}|^2 + |\tilde{\nabla} \tilde{\Phi}|^2 + \frac{\tilde{\lambda}}{24} (\tilde{\chi}^2-1)^2 + \tilde{\Lambda}\tilde{\chi}|\tilde{\Phi}|^2 + \tilde{\chi}^2 |\tilde{\Phi}|^2 \right] = \frac{\sqrt{\kappa}}{v} E.
\end{equation}
Note that several important ratios are the same as for physical and dimensionless ones, e.g. $\tilde{E}/\tilde{Q}\tilde{m}_\Phi = E/Qm_\Phi$, and $\tilde{\omega}/\tilde{m}_\Phi = \omega/m_\Phi$.
Although we find it clearer to use the original dimensionful parameters in the analytical analysis of this section, the dimensionless parameters are well-suited for numerical analysis and are shown in the figures in section.~\ref{ssec:numericalresults}.

\subsection{Classical and quantum Stability}
Before describing the details of the Q-ball solutions, we define two stability conditions.
First, classical stability states that the solution is stable (i.e. will not be dissipated) under arbitrary perturbative deformations \textit{as long as quantum effects are ignored}.
Mathematically, this means that the solution must be a local minimum of the energy functional eq.~\eqref{FLS Hamiltonian}
under the constraint $Q=\mathrm{const.}$, which is expressed as an inequality for a second-order variation of the energy,
\begin{align}
\text{classical stability:} \quad  (\delta^2 E)_Q \geq 0
\label{cls-stability}
\end{align}
for arbitrary field variations $\delta \phi$, $\delta \chi$ around the solution.
Due to the constraint, this problem cannot be solved by a simple eigenvalue analysis of Hessian matrix ($\delta^2 E/\delta \phi^2$ etc.) and requires non-trivial considerations.
Based on the theorems given in ref.~\cite{Friedberg:1976me}, we present necessary and sufficient conditions for the classical stability eq.~\eqref{thm-stability} in appendix~\ref{app:stab}, which includes the stability against fission \cite{Lee:1991ax, Tsumagari:2008bv, Mai:2012yc}.

The second kind of stability is quantum stability, which states that the solution is stable against quantum tunneling processes. This requires the solution to be the global minimum of the energy functional. 
As shown below, there is at most one Q-ball solution that is classically stable for given parameters for our model, while a plane-wave solution consisting of homogeneously distributed free particles is another local minimum. Although the condition $\omega < m_\Phi$ for Q-balls prohibits the emission of free particles as $dE/dQ = \omega$ (eq.~\eqref{eq:dEdQ}) \cite{Gulamov:2013cra}, it does not restrict the total decay into free particles. Thus, a classically stable Q-ball solution is a global minimum if it has a lower energy than the plane-wave solution, so
\begin{align}
\text{quantum stability:} \quad E \leq m_\Phi^{} Q 
\label{qtm-stability}
\end{align}
where the right-hand side of the inequality corresponds to the energy of $Q$ quanta of $\Phi$. 
Q-balls satisfying this condition are absolutely stable and cannot decay, while only classically stable but not quantum stable ones decay by tunneling effects and have finite lifetimes, although the probabilities are exponentially suppressed. Solutions of the latter are called meta-stable solutions.

\begin{figure}[t]
\centering
\includegraphics[width=0.49\textwidth]{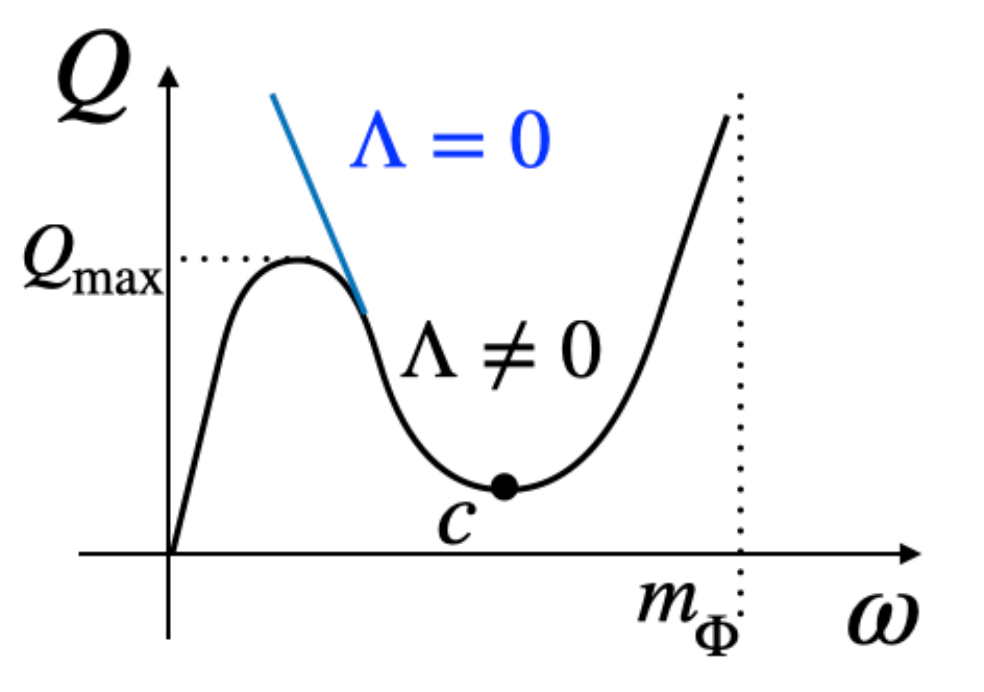}
\includegraphics[width=0.49\textwidth]{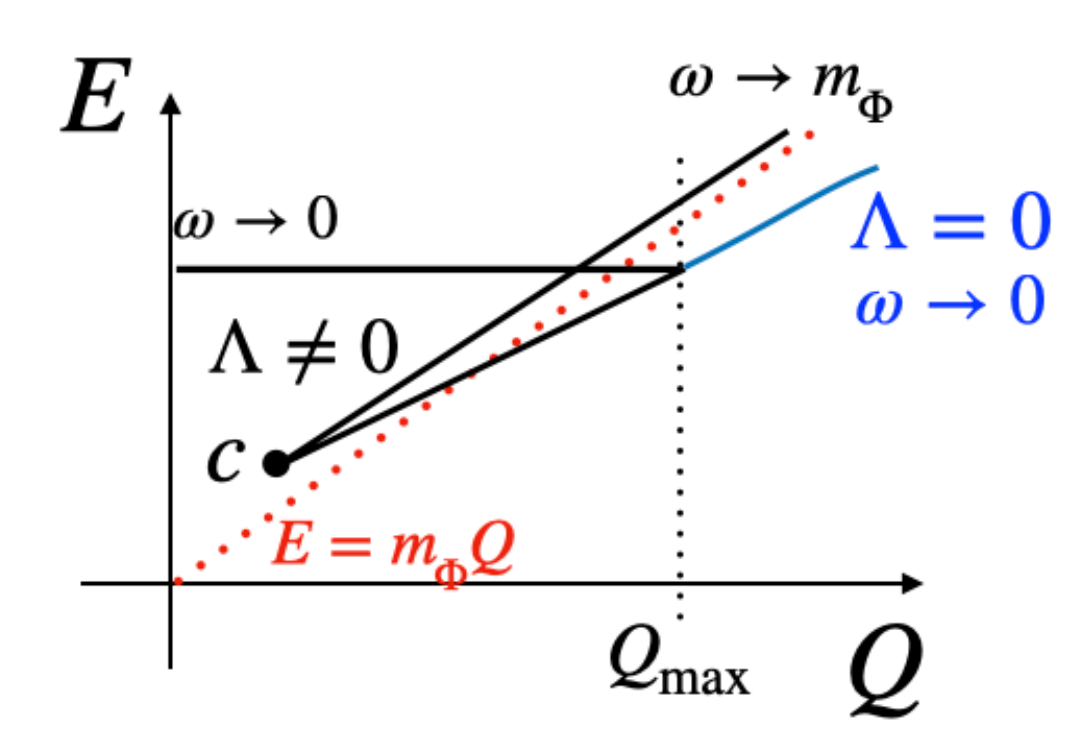}
\caption{
Left (Right): qualitative behavior of $Q(\omega)~(E(Q))$. 
The black (blue) contour corresponds to $\Lambda \neq 0~(=0)$. 
When $\Lambda\neq 0$, there exists a maximum value of $Q$.
}
\label{fig:Q-omega}
\end{figure}

In figure~\ref{fig:Q-omega}, we show the qualitative behaviors of $Q(\omega)$ (left) and $E(Q)$ (right).
%
For $\Lambda\neq0$ (black), there can be three solutions with different $\omega$ for the same $Q$, which belong to different branches of Q-ball. 
The small and large $\omega$ branches have $dQ/d\omega > 0$ so they are physically unstable and do not satisfy the condition of classical stability in eq.~\eqref{thm-stability}. 
Therefore, while $\Lambda=0$ (blue) has classically stable Q-ball solutions for $Q>Q_{\rm min}$, $\Lambda\neq 0$ (black) has a maximum charge $Q_\mathrm{max}$ in addition to $Q_{\rm min}$.
See appendix~\ref{app:stab} for the details of the classical stability. 
And in the right panel, we see that classically stable solutions with very small $Q$ near $Q_{\rm min}$ fail to satisfy the quantum stability condition.
Note also that $\omega\rightarrow m_\Phi^{}$ corresponds to the plane-wave solution as explained before.

\subsection{Thin-wall solutions}\label{ssec:thin-wall}

The Q-ball solution can be found in principle by directly solving eqs.~\eqref{eq:QballEOM}.  
This corresponds to scanning the whole space of spherically symmetric stationary field configurations given by eq.~\eqref{eq:Qballansatz}. 
However, for certain limits of the parameter space, we can obtain approximate analytical solutions and understand the behaviors of Q-balls. 
This happens in the thin- and the thick-wall limits, as will be shown below. 

The thin-wall approximation holds when $\phi$ inside the Q-ball is large enough so that $\chi$ varies only near the Q-ball boundary where $\phi$ inevitably goes to zero. 
This can be captured by eqs.~\eqref{eq:QballEOM}, that when $\phi$ is too large, $\chi$ should be nearly fixed to $\chi = -\Lambda / 2 \kappa$ to make the two terms with $\phi^2$ cancel each other. 
This constant $\chi$ gives a sinc function solution to $\phi$ via eqs.~\eqref{eq:QballEOM}, with the Q-ball radius given by its first root. 
Beyond this point, once $\phi$ becomes small, $\chi$ quickly goes to its VEV at $\chi = v$, and $\phi$ follows the trivial solution at $\phi = 0$. 

A large enough value of $\phi$ corresponds to a large localized charge. In this case, the spatial size of the Q-ball is large, so that the vacuum energy in the inside region is dominant over the surface energy from the variation of $\chi$ in Eq.~\eqref{modified energy} for $\Lambda/2\kappa \ll 1$. 
By completely neglecting the wall region, the solution can be approximated to the lowest order as 
\begin{eqnarray}
    \chi(r) \simeq 
    \begin{cases}
		-\frac{\Lambda}{2\kappa} & r < R_\Lambda \\
        v & r > R_\Lambda
    \end{cases}
    \, , \qquad
    \phi(r) \simeq
    \begin{cases}
        \phi_0 \frac{\sin \omega_\Lambda r}{\omega_\Lambda r} & r < R_\Lambda \\
        0 & r > R_\Lambda
    \end{cases} \label{eq:thinwall},
\end{eqnarray}
where $R_\Lambda = \pi/\omega_\Lambda$ with $\omega_\Lambda \equiv (\omega^2 + \Lambda^2/4\kappa)^{1/2}$. 
Substituting this approximate solution eq.~\eqref{eq:thinwall} into eq.~\eqref{eq:Q}, the charge of the Q-ball is found to be
\begin{equation}
    Q \simeq \frac{2\pi^2 \phi_0^2 \omega}{\omega_\Lambda^3}. \label{eq:Qthinwall}
\end{equation}
Since eq.~\eqref{eq:Qthinwall} relates the two parameters $\omega$ and $\phi_0$ for a given $Q$, the thin-wall configuration is fully described by the single parameter $\omega$. The value of $\omega$ that represents the Q-ball solution is found by minimizing the energy with respect to $\omega$, which then fixes $\phi_0$ through eq.~\eqref{eq:Qthinwall}.

Eq.~\eqref{eq:thinwall} is discontinuous for $\chi$ and $\phi'$ at $r = R_\Lambda^{}$, the Q-ball radius. Indeed, more sophisticated solutions having continuous fields and derivatives can be obtained by imposing smooth junction conditions at $r = R_\Lambda$. For example, ref.~\cite{Friedberg:1976me} imposes continuity for $\chi$, while ref.~\cite{Heeck:2023idx} imposes continuity for $\phi'$. However, the resultant solutions are still a piecewise approximation, and eq.~\eqref{eq:thinwall} is sufficient for our purpose since the wall's contribution to the total energy is negligible in the thin-wall limit; we examine the thick-wall regime in the next subsection. In section~\ref{ssec:numericalresults}, we will see that the approximate solution eq.~\eqref{eq:thinwall} agrees well with numerical results in the large $Q$ and small $\Lambda$ regime.

Within the thin-wall approximation, the total energy is evaluated as 
\aln{
    E & = \int d^3x \left( \frac{1}{2} \left( \frac{d\chi}{dr} \right)^2 + \frac{1}{2} \omega^2 \phi^2 + \frac{1}{2} \left( \frac{d\phi}{dr} \right)^2 + U(\chi) + \frac{1}{2} \Lambda \chi \phi^2 + \frac{1}{2} \kappa \chi^2 \phi^2 \right)\label{eq:energyintegral}
   \\
     &\simeq \frac{2 \pi^2 \omega^2}{\omega_\Lambda^3} \phi_0^2 + \frac{\lambda \pi^4}{18\omega_\Lambda^3} \left(\frac{\Lambda^2}{4\kappa^2} - v^2 \right)^2 = \omega Q + \frac{\lambda \pi^4}{18\omega_\Lambda^3} \left(\frac{\Lambda^2}{4\kappa^2} - v^2 \right)^2. \label{eq:Ethinwall}
}
Differentiating eq.~\eqref{eq:Ethinwall} by $\omega$ gives
\begin{equation}
    \frac{dE}{d\omega} = Q - \frac{\lambda \pi^4 \omega}{6 \left( \omega^2 + \frac{\Lambda^2}{4\kappa} \right)^{5/2}} \left(\frac{\Lambda^2}{4\kappa^2} - v^2 \right)^2~,
    \label{eq:dEdomegathinwall}
\end{equation}
and the thin-wall solution $\omega=\omega_{\rm sol}^{}$ is obtained by solving $dE/d\omega = 0$ with $d^2E/d\omega^2 > 0$. Eq.~\eqref{eq:dEdomegathinwall} shows an important difference between $\Lambda = 0$ and $\Lambda \neq 0$. 
When $\Lambda = 0$, $E(\omega)$ always has a local minimum for $\omega > 0$. 
However, when $\Lambda \neq 0$, the second term in  eq.~\eqref{eq:dEdomegathinwall} is bounded from above, indicating that $dE/d\omega$ can never be $0$ for $Q$ exceeding the maximum value and hence the Q-ball solution does not exist. 
The second term in eq.~\eqref{eq:dEdomegathinwall} has its maximum at $\omega = |\Lambda|/4\sqrt{\kappa}$ (restricting to $\omega > 0$), giving the maximum $Q$ for $\Lambda \neq 0$ in the thin-wall regime as
\begin{equation}
    Q^{\rm thin}_{\rm max} = \frac{128 \pi^4 \kappa^2 \lambda}{75\sqrt{5} \Lambda^4}  \left(\frac{\Lambda^2}{4\kappa^2} - v^2 \right)^2~,
    \label{eq:Qmaxthinwall}
\end{equation}
and the corresponding energy is 
\begin{equation}
    E\simeq 0.67\frac{\Lambda}{\sqrt{\kappa}}Q_{\rm max}^{\rm thin}~,
\end{equation}
which can be compared to the plane-wave case $m_{\Phi} Q_{\rm max} = \sqrt{\Lambda v +\kappa v^2} \, Q_{\rm max}$ for the quantum stability of the Q-ball.

\subsection{Thick-wall solutions}
\label{ssec:thickwall}
We now consider the thick-wall regime where the energy contribution from the variation of $\chi$ in the boundary region is dominant over the vacuum energy in the bulk interior. 
To take into account the significant width of the wall, we now consider the continuous solution in $\chi$ used in ref.~\cite{Friedberg:1976me}:
\begin{eqnarray}
    \chi(r) \simeq 
    \begin{cases}
		-\frac{\Lambda}{2\kappa} & r < R_\Lambda \\
        -\frac{\Lambda}{2\kappa}+\left(v+\frac{\Lambda}{2\kappa}\right)\left[1-e^{-m_\chi(r-R_\Lambda)}\right] & r > R_\Lambda
    \end{cases}
    \, , \qquad
    \phi(r) \simeq
    \begin{cases}
        \phi_0 \frac{\sin \omega_\Lambda r}{\omega_\Lambda r} & r < R_\Lambda \\
        0 & r > R_\Lambda
    \end{cases}. \nonumber \\ \label{eq:thickwall}
\end{eqnarray}
This approximate solution is valid for $Q$ large enough to keep $\chi\simeq -\Lambda/2\kappa$ inside the Q-ball such that the last two terms in eqs.~\eqref{eq:QballEOM} dominate the EOM for $\chi$. 

By integrating the $(d\chi/dr)^2$ term in eq.~\eqref{eq:energyintegral} for $r > R_\Lambda^{}$\footnote{We omit the additional vacuum energy contribution in $U(\chi)$ that results from the exponential behavior of $\chi$ for $r>R_\Lambda^{}$. 
}, we find the surface energy to be composed of three terms with different $\omega_\Lambda$ scalings,
\begin{equation}
\label{eq:esurf}
    E_{\rm surf} = \frac{2\pi^3 v^2 m_\chi}{\omega_\Lambda^2}a_1 + \frac{4\pi^2 v^2}{\omega_\Lambda}a_2 + \frac{4\pi v^2}{m_\chi}a_3~,
\end{equation}
where
\begin{align}
\begin{split}
\label{eq:surfcoeff}
    a_1=& \left(1+\frac{\Lambda}{2\kappa v}\right)^2 -\frac{1}{3}\left(1+\frac{\Lambda}{2\kappa v}\right)^3 + \frac{1}{16}\left(1+\frac{\Lambda}{2\kappa v}\right)^4~, \\
    a_2=& \frac{1}{2}\left(1+\frac{\Lambda}{2\kappa v}\right)^2 -\frac{1}{9}\left(1+\frac{\Lambda}{2\kappa v}\right)^3 + \frac{1}{64}\left(1+\frac{\Lambda}{2\kappa v}\right)^4~, \\
    a_3=& \frac{1}{4}\left(1+\frac{\Lambda}{2\kappa v}\right)^2 -\frac{1}{27}\left(1+\frac{\Lambda}{2\kappa v}\right)^3 + \frac{1}{256}\left(1+\frac{\Lambda}{2\kappa v}\right)^4~.   
\end{split}
\end{align}
The surface energy contributions are clearly different for positive and negative $\Lambda$, and increase with increasing $\Lambda$. 

Now let us evaluate the maximum value of $Q$ in the thick-wall regime. 
We can minimize the effective energy by taking the derivative with respect to $\omega$,
\begin{equation}
\label{eq:dedwtotal}
    \frac{dE}{d\omega} = Q - \frac{\lambda \pi^4 \omega}{6\omega_\Lambda^5}\left(\frac{\Lambda^2}{4\kappa^2}-v^2\right)^2 - \frac{4\pi^3 v^2 m_\chi \omega}{\omega_\Lambda^4} a_1 - \frac{4\pi^2 v^2 \omega}{\omega_\Lambda^3} a_2 = 0~.
\end{equation}
From the second derivative, the maximum of the second, third, or fourth terms individually is found to be at
\aln{
\omega = \frac{\Lambda}{4\sqrt{\kappa}}~,\quad  \frac{\Lambda}{2\sqrt{3\kappa}}~,\quad  \frac{\Lambda}{2\sqrt{2\kappa}}~
} 
respectively, so all three terms can be maximized with similar values of $\omega$. 
Near the limiting value of $Q\rightarrow Q_{\rm max}$, i.e. $\omega_\Lambda={\cal O}(\Lambda/\sqrt{\kappa})$, one can check that the first term of the $a_1$ contribution is dominant in eq.~(\ref{eq:esurf}) for
\aln{
 0\leq \left|\frac{\Lambda}{2\kappa v}\right|\lesssim 1~,\quad \frac{\sqrt{\lambda}}{\kappa}\sim{\cal O}(1)~.
}

Since the largest surface energy contribution at moderate values of $\Lambda$ is the $a_1$ term, we set $\omega=\tfrac{\Lambda}{2\sqrt{3\kappa}}$ in our analysis. 
Comparing eqs.~\eqref{eq:Ethinwall} and~\eqref{eq:esurf}, the volumetric vacuum energy term is dominant for $\Lambda/2\kappa v \lesssim 1/4$, justifying the thin-wall approximation for small $\Lambda$. In the thick-wall regime, the $a_1$ term is dominant for $1/4 \lesssim \Lambda/2\kappa v \lesssim 2\pi$, and the $a_2$ term is dominant for $\Lambda/2\kappa v \gtrsim 2\pi$.

We now calculate $Q_{\rm max}$ using only the contribution from the $a_1$ surface term in combination with the $Q$ term. Comparing the coefficients of this surface term in eq.~\eqref{eq:surfcoeff}, the $(1+\Lambda/2\kappa v)^2$ term dominates up to $\Lambda/2\kappa v \leq 2$. Therefore, the maximum charge allowed in the thick-wall approximation for $1/4<\Lambda/2\kappa v<2$ is 
\begin{equation}
\label{eq:qmaxsurf}
    Q^{\rm thick}_{\rm max} = \frac{6\sqrt{3}\pi^3 v^2 m_\chi \kappa^{3/2}}{\Lambda^3} a_1 \approx \frac{6\sqrt{3}\pi^3 v^2 m_\chi \kappa^{3/2}}{\Lambda^3} \left(1+\frac{\Lambda}{2\kappa v}\right)^2~.
\end{equation}
We show this line in figure~\ref{fig:LambdaQplane}, where it roughly reproduces the scalings $Q_{\rm max} \sim \Lambda^{-3}$ for $1/4<\Lambda/2\kappa v <1$ and $Q_{\rm max} \sim \Lambda^{-1}$ for $1<\Lambda/2\kappa v < 2$.  
It can be seen from eq.~\eqref{eq:surfcoeff} that the surface energy contributions with $\Lambda>0$ are much larger than those with $\Lambda<0$ so that negative $\Lambda$ solutions are well-approximated by the thin wall. 

It could be interesting to consider the limiting case of zero scalar $\chi$ potential, $\lambda\xrightarrow{} 0$, as discussed in e.g. ref.~\cite{Levin:2010gp}. In this regime, the thick-wall approximation would hold and from eq.~\eqref{eq:qmaxsurf}, we naively expect the maximum charge to become vanishingly small. Our analysis is based off of eq.~\eqref{eq:thickwall}, but for $\lambda\xrightarrow{} 0$, the exponential in that equation should be replaced by $1/r$.

\section{Numerical analysis} \label{sec:numerical}
Our numerical analysis broadly confirms the analytic treatment in the previous section, while providing more detailed and accurate results. Here, we implement two complementary methods to solve the Q-ball EOMs~\eqref{eq:QballEOM}, the traditional non-linear Richardson iteration~\cite{kelley1995iterative} and the modified gradient flow method~\cite{Chigusa:2019wxb}. In the following, we briefly explain these two methods and compare their convergence conditions. 

There are two different formalisms for obtaining the Q-ball solutions.
The first solves for stationary configurations of the energy functional $E$ with charge $Q$ fixed, which satisfies Variational Principle 1 presented in appendix~\ref{app:stab}.
In this formalism, $\omega$ is not an independent parameter but depends on the field $\phi$ and $Q$. 
Alternatively, $\omega$ can be regarded as the input parameter instead of $Q$. 
From this viewpoint, the conservation law for $Q$ is not taken into account when solving the EOMs and $Q$ is determined from the obtained solution $\phi$ and input parameter $\omega$. 
This formalism satisfies Variational Principle 2 in appendix~\ref{app:stab}.
The nonlinear Richardson iteration and the modified gradient flow method correspond to Variational Principles 1 and 2, respectively.

\subsection{Numerical methods} \label{ssec:numericalmethods}

\noindent{\bf Nonlinear Richardson iteration.} The nonlinear Richardson iteration is a fixed point iteration that obtains the solution to the EOMs through Variational Principle 1, in which a nonlinear equation $K[f(x)] = 0$ is solved by a successive approximation as $f_{n+1} (x) = f_n(x) + \epsilon K[f_n(x)]$, where $f_n(x)$ is the $n$-th trial solution and $\epsilon$ is a constant parameter that can be chosen empirically. It is obvious that the iteration has a fixed point at the true solution since if $K[f_n(x)] = 0$, further iterations give no change to the trial solution. From our Q-ball EOMs~\eqref{eq:QballEOM}, the iteration equations for $\chi(r)$ and $\phi(r)$ are
\begin{subequations}
\begin{eqnarray}
    \chi_{n+1} &=& \chi_n + \epsilon \, \left[\frac{1}{r^2} \frac{d}{dr} \left(r^2 \frac{d\chi_n}{dr} \right) - \frac{dU(\chi_n)}{d\chi_n} - \frac{\Lambda}{2} \phi_n^2 - \kappa \phi_n^2 \chi_n \right], \\
    \phi_{n+1} &=& \phi_n + \epsilon \, \left[\frac{1}{r^2} \frac{d}{dr} \left(r^2 \frac{d\phi_n}{dr} \right) + \omega^2 \phi_n - (\Lambda \chi_n + \kappa \chi_n^2) \phi_n \right].
\end{eqnarray}%
\label{eq:Richardson}%
\end{subequations}

The numerical evaluation of r.h.s. of eqs.~\eqref{eq:Richardson} is done by first compactifying the semi-infinite $r$ domain into $0 \leq s \leq 1$ by $r/(R_0 + r)$ for an empirically chosen $R_0$ (similar to \cite{Heeck:2020bau, Heeck:2023idx}), then making a regular grid on $s$ domain, and finally approximating the derivatives with finite differences. The boundary conditions for $\chi$ and $\phi$ are $\chi'(s=0) = 0$, $\phi'(s=0) = 0$, $\chi(s=1) = v$, and $\phi(s = 1) = 0$. A tricky part of the iteration is that the change of $\phi$ at each step does not conserve $Q$, given by the integration in eq.~\eqref{eq:Q}. To find the solution of a given $Q$, we thus adjust $\omega$ at each step to make $Q$ a constant during the iteration. 

To see when the iteration converges, we assume an $n$-th trial solution in the vicinity of the true solution $f_{\rm sol}(x)$ as $f_n(x) = f_\mathrm{sol}(x) + \delta f_n(x)$, with $|\delta f_n(x)|\ll 1$. Then, the $(n+1)$-th trial solution is determined by
\begin{align}
&  f_\mathrm{sol}(x) + \delta f_{n+1}(x) = f_\mathrm{sol}(x) + \delta f_{n}(x) + \epsilon K[f_\mathrm{sol}(x)+ \delta f_{n}(x)]\\
\therefore  \, & \delta f_{n+1}(x) = \delta f_{n}(x) - 
\epsilon \int d^3 y \, \left.\left(\frac{\delta^2 E}{\delta f(x) \delta f(y)} \right)_Q \,\right|_{f=f_\mathrm{sol}} \delta f_{n}(y)+\mathcal{O}(\delta f ^2)
\end{align}
where we have used that $K[f_\mathrm{sol}]$ vanishes and $K$ is given by (minus of) the first-order derivative of the energy functional with fixed $Q$, $K=-(\delta E / \delta f)_Q$.
From this, it follows that the condition of the convergence $|\delta f_{n+1}/\delta f_n|<1$ is stated by that the second-order variation $(\delta^2 E)_Q$ is positive for arbitrary variation $\delta f_{n}(x)$. 
This is precisely the classical stability condition for the Q-ball solution eq.~\eqref{cls-stability}.
\\

\noindent{\bf Modified gradient flow.} Let us move on to the other method, the modified gradient flow in the Variational Principle 2,
in which the functional $\mathcal{E}$ defined in eq.~\eqref{modified energy} is regarded as a Legendre transform of the energy functional $E$, 
and hence the solutions of the EOMs~\eqref{eq:QballEOM} are stationary points of $\mathcal{E}$ with an independent parameter $\omega$ fixed through the calculation (see appendix~\ref{app:stab} for the details).
Due to Derrick's theorem (Theorem 1 in appendix~\ref{app:stab}),
this stationary point must be a saddle point of $\mathcal{E}$ instead of a local minimum and have an unstable mode around it (i.e., the second-order curvature in that direction is negative).
Thus a naive gradient flow, which is also known as the steepest descent flow,
\begin{align}
    \chi_{n+1} &= \chi_n -\epsilon \frac{\delta \mathcal{E}[\chi_n,\phi_n]}{\delta \chi_n} \\
    \phi_{n+1} &= \phi_n -\epsilon \frac{\delta \mathcal{E}[\chi_n,\phi_n]}{\delta \phi_n} 
\end{align}
fails to converge to the solutions. This situation is very similar to those of bounce solutions~\cite{Coleman:1977py}.

The modified gradient flow method is introduced to obtain bounce solutions in ref.~\cite{Chigusa:2019wxb} and applied to some models in refs.~\cite{Ho:2019ads,Hamada:2020rnp,Ho:2020ltr}.
It is able to obtain them successfully by adding appropriate ``modification terms'' in the flow equations as
\begin{align}
    \chi_{n+1} &= \chi_n -\epsilon \frac{\delta \mathcal{E}[\chi_n,\phi_n]}{\delta \chi_n}  + \epsilon \beta \mathcal{G}_\chi[\chi_n,\phi_n]\\
    \phi_{n+1} &= \phi_n -\epsilon \frac{\delta \mathcal{E}[\chi_n,\phi_n]}{\delta \phi_n}  + \epsilon \beta \mathcal{G}_\phi[\chi_n,\phi_n]
\end{align}
with a constant $\beta>1$ and 
\begin{align}
   \mathcal{G}_\chi[\chi_n,\phi_n] &= \partial_r \chi_n 
   \left(\int dr' \, \frac{\delta \mathcal{E}[\chi_n,\phi_n]}{\delta \chi_n(r')} \partial_{r'} \chi_n (r')\right)  
   \left(\int dr'' \, (\partial_{r''} \chi_n(r''))^2\right)^{-1}\\
   \mathcal{G}_\phi[\chi_n,\phi_n] &= \partial_r \phi_n
   \left(\int dr' \, \frac{\delta \mathcal{E}[\chi_n,\phi_n]}{\delta \phi_n(r') } \partial_{r'} \phi_n (r')\right)  
   \left(\int dr'' \, (\partial_{r''} \phi_n (r''))^2\right)^{-1} \, ,
\end{align}
where we have chosen the modification terms as the same as ref.~\cite{Chigusa:2019wxb}.
The charge $Q$ has nothing to do in the EOMs since it is explicitly written in terms of $\omega$ and $\phi$, and hence it is calculated by substituting the fields and $\omega$ into eq.~\eqref{eq:Q} once the solutions are obtained.
We naively discretize the spatial coordinate $r$ and put the same boundary conditions as those in the nonlinear Richardson iteration above.

This modified gradient flow method converges if and only if the solution has only one negative mode (unstable mode) under the original gradient flow without modification, or equivalently, if and only if the Hessian matrix $H$ in appendix~\ref{app:stab} has only one negative eigenmode.
According to the theorem~\eqref{thm-stability}, this is a necessary condition for the classical stability, and hence is loose compared to the convergence condition in the nonlinear Richardson iteration.
Indeed, not all solutions found in the modified gradient flow are obtained in the nonlinear Richardson iteration.
Among the found solutions, those satisfying $dQ(\omega) / d\omega < 0$ are physically stable (see Theorem 3 in appendix.~\ref{app:stab}) and can also be obtained by the nonlinear Richardson iteration.
\\

\subsection{Results} \label{ssec:numericalresults}
We now present the numerical results. We use dimensionless variables defined in section~\ref{ssec:Lagrangian} in plots, and use $\tilde{\lambda} = 1$ as a reference case.
\begin{figure}[tbp] \centering
\includegraphics[width=0.49\linewidth]{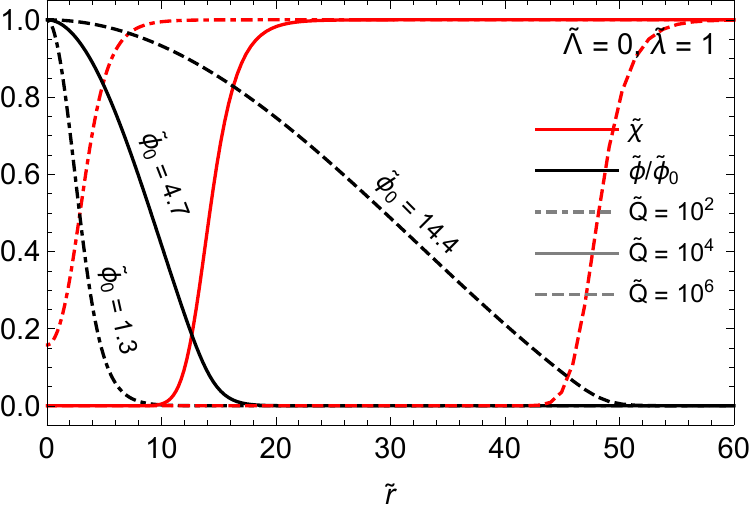}
\includegraphics[width=0.49\linewidth]{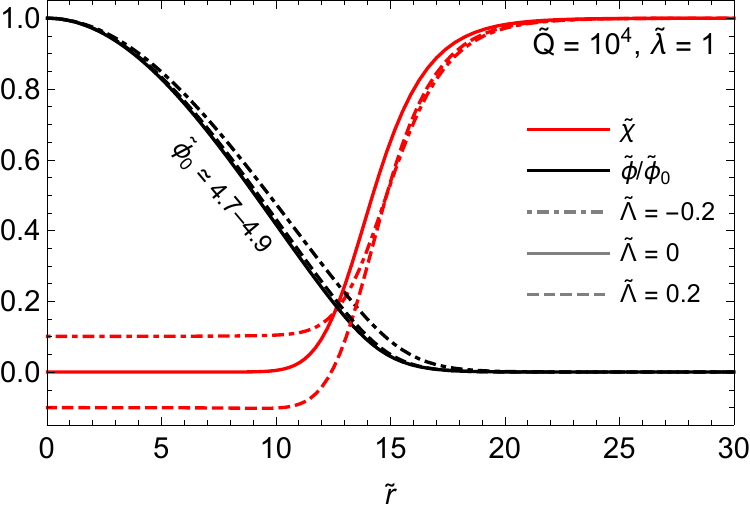}
\caption{Radial field profiles for Q-ball for varying $Q$ with $\Lambda$ being fixed (left) and varying $\Lambda$ with $Q$ being fixed (right). The dimensionless fields $\tilde{\chi} = \chi / v$ (red) and $\tilde{\phi} = \phi / v$ (black) are plotted as functions of dimensionless radius $\tilde{r} = \sqrt{\kappa} v r$. $\tilde{\phi}$ is normalized by its value at $r = 0$.
} 
\label{fig:fieldprofile}
\end{figure}
%
Figure~\ref{fig:fieldprofile} shows the radial field profiles for selected example Q-ball configurations where $\tilde{\phi}(r)/\tilde{\phi}(0)$ and $\tilde{\chi}(r)$ are drawn by black and red curves, respectively. For all the cases, $\phi$ has localized non-zero value in a region of some finite radius, where $\chi$ inside also departs from its true vacuum, hence being a non-topological soliton. 

In the left panel, we vary $Q$ with $\Lambda$ being fixed to be zero. 
As $Q$ increases, we see that both the Q-ball radius and $\phi$ inside increase, which is physically expected as the localized charge becomes greater. 
For large $Q$'s, $\chi$ almost maintains a constant value inside, and the variation happens rapidly near the boundary of the Q-ball. 
While the absolute width of this wall does not change much, its relative width with respect to the overall Q-ball radius decreases as $Q$ increases. 
The fraction of the energy residing in the wall also decreases and becomes negligible, which results in the thin-wall limit. 
On the other hand, for small $Q$,  we see that $\chi$'s variation happens continuously in the entire Q-ball region with $\phi \neq 0$.

In the right panel, we vary $\Lambda$ with the charge being fixed at $\tilde{Q}=10^4$. 
Once the total charge is fixed, we see that $\phi$'s profile does not change much, while $\chi$ inside changes according to $\chi \simeq -\Lambda / 2 \kappa$ as discussed in section~\ref{ssec:thin-wall}. 
This is true only for large $Q$ cases with large $\phi$ inside, as we see for the lowest $Q$ case in the left panel in which $\chi$ takes a different value $\neq -\Lambda/2\kappa=0$ around the center of the Q-ball. This is because when $\phi$ becomes small, the gradient term of $\chi$ is no longer negligible in the total energy and is subject to minimization.

\begin{figure}[tbp]
\centering
\includegraphics[width=0.48\textwidth]{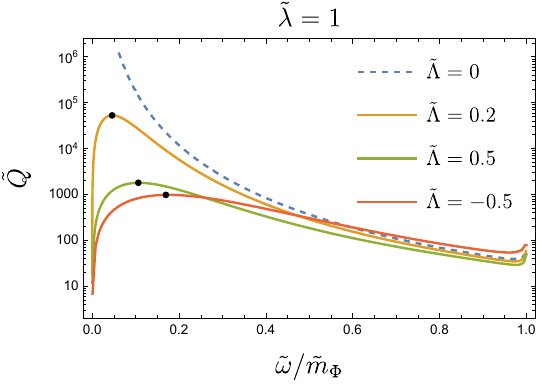}
\hspace{1em}
\includegraphics[width=0.48\textwidth]{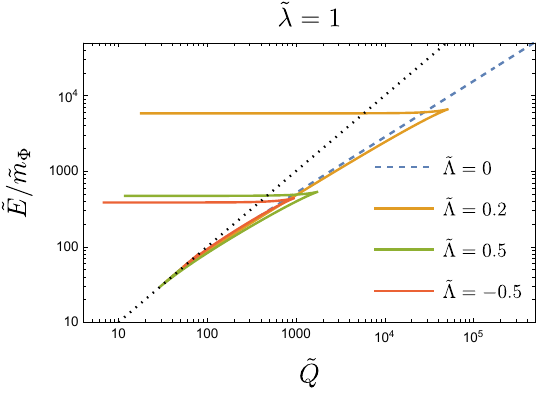}
\caption{Plots of curves of $\tilde{Q}$ vs $\tilde{\omega}/\tilde {m}_\Phi$ (left) and $\tilde{E}/\tilde{m}_\Phi$ vs $\tilde{Q}$ (right) for solutions obtained by the modified gradient flow method with $\tilde{\Lambda}=0, 0.2$ and $\pm 0.5$. Left: the blue dashed line ($\tilde \Lambda=0$) diverges for $\tilde\omega\to 0$ as there is no maximum value of $Q$ whereas the other three lines with $\tilde \Lambda\neq 0$ approach $0$ since $Q$ is proportional to $\omega$.
All of them decrease as $\tilde\omega$ increases in the intermediate regime, and suddenly turn to increase around $\tilde{\omega} \to \tilde{m}_\Phi$, which corresponds to the plane wave solution.
The black dots indicate the maximum points of the curves.
Note that only the solutions with decreasing $\tilde{Q}(\tilde{\omega})$ are physically stable, see the text.
Right: the blue dashed line with $\tilde \Lambda=0$ can have infinitely large $\tilde{Q}$ and $\tilde{E}$, while the other three lines turn left at finite $\tilde{Q}$. These correspond to the behaviors for $\tilde \omega\to 0$ in the left panel.
The black dotted line indicates $\tilde{E} / \tilde{Q}\tilde{m}_\Phi = E / Q m_\Phi = 1$.
}
\label{fig:Q-E-omega}
\end{figure}

In figure~\ref{fig:Q-E-omega}, we scan the parameter space and show relations between $\omega$, $Q$, and $E$. The modified gradient flow method is used to explore all the Q-ball solutions, regardless of the stability.
For a given set of parameters appearing in the Lagrangian, selecting $\omega$ fixes the Q-ball profile, and $Q(\omega)$ is 
shown in the left panel. 
For the obtained solutions, we show the relation between $Q$ and $E$ in the right panel. 
As we discussed with figure~\ref{fig:Q-omega}, the solutions with $\Lambda\neq 0$ show different behaviors around $\omega \to 0$ compared to the case of $\Lambda=0$. 

In the left panel, the charge $Q$ diverges as $\omega \to 0$ for $\Lambda=0$ while it suddenly approaches $0$ as $\omega \to 0$ for $\Lambda\neq 0$. 
This behavior can be understood from eq.~\eqref{eq:Qthinwall}, where $\omega_\Lambda=\sqrt{\omega^2+ \Lambda^2/4\kappa}$ does not vanish even if $\omega$ vanishes, due to the non-zero $\Lambda$.
For all four cases, the charge decreases as $\omega$ increases in the intermediate regime, and suddenly turns to increase around $\omega \to m_\Phi$ corresponding to the plane wave solution.
These behaviors are consistent with the qualitative plots shown in figure~\ref{fig:Q-omega}.  
The black dots indicate the maximum points of the curves,
from which one can see that the minimum value of $\omega$ to realize the stable Q-balls gets larger as $|\Lambda|$ increases.
This is consistent with the prediction made in the thin-wall approximation that $Q$ is maximized at $\omega = |\Lambda|/(4\sqrt{\kappa})$
(see the text below eq.~\eqref{eq:dEdomegathinwall}).
Note that the minimum values of $\omega$ for the stability are almost insensitive to the sign of $\Lambda$
although the normalized values divided by $\tilde{m}_\Phi$ become different as shown in the left panel ($|\tilde\Lambda|=0.5)$.

In the right panel, the solutions with $\Lambda=0$ can have infinitely large $Q$ and $E$, corresponding to $\omega\to 0$.
On the other hand, the other three lines with $\Lambda \neq 0$ suddenly turn at around $\tilde{Q}\sim 10^3$ to left being almost horizontal, corresponding to $\omega \to 0$.
This branch is not classically stable because it corresponds to the $dQ/d\omega>0$ branch in the left panel.
The black dotted line indicates $E=Q m_\Phi$, below which classically stable solutions satisfy the quantum stability~\eqref{qtm-stability}.
These behaviors can also be seen in figure~\ref{fig:Q-omega}.  
Note that it is difficult to obtain solutions in the region with $\omega \to m_\Phi$ by the modified gradient flow method since the flow does not converge due to the appearance of the second negative mode, as stated above.
Thus the branch beyond the point $c$ in figure~\ref{fig:Q-omega} is only partially reproduced.  
Nevertheless, such a branch is not classically stable by Theorem 3 in appendix~\ref{app:stab}, and hence is not significant in our phenomenological/cosmological argument.

\begin{figure}[tbp] \centering
\includegraphics[width=0.98\linewidth]{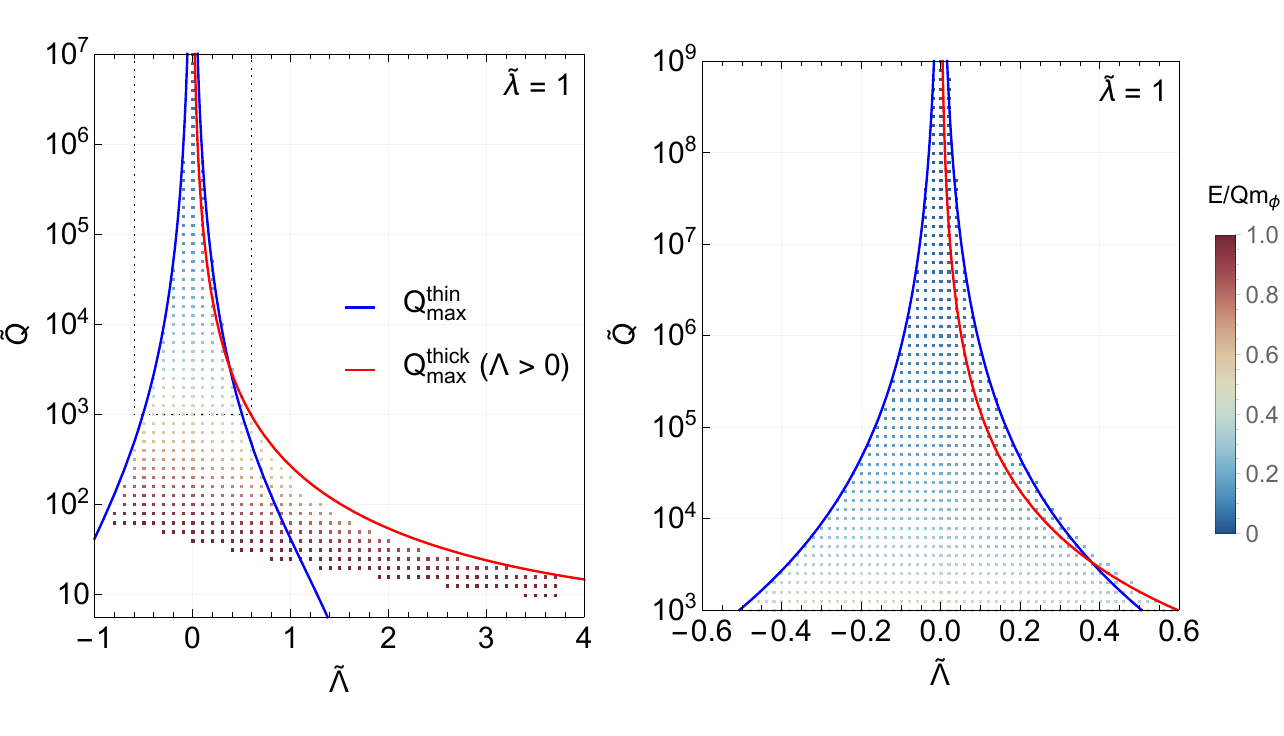}
\caption{Stable Q-ball solutions plotted on the $\tilde{\Lambda}$-$\tilde{Q}$ plane with $\tilde{\lambda} = 1$. $E/Qm_\Phi$ is displayed with colors. Maximum charges predicted from the thin-wall approximation (eq.~\eqref{eq:Qmaxthinwall}) and the thick-wall approximation (eq.~\eqref{eq:qmaxsurf}) are overlaid. The entire parameter space for each plot is scanned with a uniform resolution. The right panel corresponds to the dashed box in the left panel, extended upward.}
\label{fig:LambdaQplane}
\end{figure}

In figure~\ref{fig:LambdaQplane}, we summarize the stable Q-ball solutions in the $\tilde{\Lambda}$-$\tilde{Q}$ plane, scanned by the nonlinear Richardson iteration. We color each solution point by $E/Qm_\Phi$, and the maximum charges predicted by the thin and thick-wall approximations are shown with solid lines. The overall behavior is presented in the left panel, while we focus on the large $Q$ region in the right one. 

The region with stable Q-ball solutions is bounded for a finite range of $\Lambda$. The lower bound on $\Lambda$ is strictly predicted to be $\tilde{\Lambda} = -1$, corresponding to an unphysical $\Phi$ mass of $m_\Phi^2 = \Lambda v + \kappa v^2 < 0$. The upper bound of $\Lambda$ is understandable by $Q^{\rm thick}_{\rm max}$, as the maximum charge allowed for the stability becomes too small for large $\Lambda$, insufficient to support the bound structure of $\Phi$ by having a non-trivial value of $\chi$ inside. Indeed, as can be inferred by $E/Qm_\Phi$, the lowest $Q$ solutions are on the verge of being free particle solutions, dissolution of Q-balls.\footnote{While the nonlinear Richardson iteration considers only the classical stability by its convergence, most of the obtained solutions are quantum stable too.
} On the contrary, large $Q$ solutions have large energy differences.

For $\Lambda$ residing in the range between the two, the stable Q-ball solutions exist only in a finite range of $Q$ with an exception at $\Lambda = 0$. These correspond to the segment of the curves $Q(\omega)$ in figures~\ref{fig:Q-omega} and \ref{fig:Q-E-omega} with $dQ/d\omega < 0$. We also see that the maximum charge predicted by the thick-wall approximation in eq.~\eqref{eq:qmaxsurf} correctly captures the overall dependence on $\Lambda$ in small $Q$ region at large $\Lambda$, while that from the thin-wall approximation in eq.~\eqref{eq:Qmaxthinwall} is highly accurate in large $Q$ region with small $\Lambda$. For the minimum charge, we provide an analytical estimation in eppendix~\ref{app:Qmin}.

The existence of a maximum value of $Q$ is a unique feature caused by the attractive interaction that can have various impacts on Q-ball formation and evolution in the early Universe. See also the discussion in the next section.

\section{Conclusions}
\label{sec:conclusions}

In this paper, we have studied FLS Q-balls with an attractive interaction $\Lambda \chi |\Phi|^2$ between a $U(1)$ charge carrying complex scalar $\Phi$ and a real scalar $\chi$ that undergoes spontaneous symmetry breaking. The attractive nature of this force has significant effects on the stability of the Q-ball, imposing a maximum stable charge $Q_{\rm max}$. In the absence of this interaction, i.e. $\Lambda=0$, analysis of the classical field equations shows the existence of a stable solution branch with a minimum charge but no maximum limit. However, in the presence of the attractive interaction, the stable solution branch terminates at a finite charge $Q_{\rm max}$ and is bounded on both ends. This upper limit steadily decreases with increasing interaction strength $\Lambda$, strongly restricting the parameter space of stable Q-ball solutions. For large values of $\Lambda$, the stable solution branch may disappear altogether, and the formation of these non-topological solitons may be forbidden. 

Our analytic study was based on the thin-wall and the thick-wall approximations, through which we solved for the Q-ball profile. The former is realized when the localized charge is large enough that the energy contribution of the wall is negligible, while the latter holds for the opposite case when the surface energy from the variation of the real scalar is dominant. We estimated the maximum charge $Q_{\rm max}$ in these two approximations by finding the limit where the local minimum in energy, representing the Q-ball solution, no longer exists. We have seen a good agreement between the analytical and numerical results.

Finally, there are possible implications for cosmology. In general, Q-balls can be produced when the Universe undergoes a phase transition in the early stages of its thermal history, and could be a dark matter candidate if the stability conditions are satisfied. In the traditional FLS Q-ball, there is no limit to the maximum charge, which means that stable Q-balls with arbitrarily large charges can be produced during a phase transition in principle~\cite{Krylov:2013qe}. However, the presence of the attractive interaction imposes a maximum charge $Q_{\rm max}^{}$ for stable Q-balls and implies that collapsing false-vacuum regions with large charges cannot form a stable configuration. These regions would likely continue shrinking while preserving the total charge due to the vacuum-energy pressure combined with the attractive interaction. These pockets could eventually collapse into primordial black holes, although an in-depth study of time-dependent field configurations is necessary to confirm this hypothesis. Thus, Q-balls with an attractive force could be another feasible production scenario of primordial black holes from cosmological phase transitions, and we leave the analysis of this scenario to future works.

\section*{Acknowledgements}
\addcontentsline{toc}{section}{Acknowledgments}
The authors thank Pyungwon Ko for many helpful discussions.
The work of K.K. is supported by KIAS Individual Grants, Grant No. 090901.   
P.L. was supported by Grant Korea NRF2019R1C1C1010050 and KIAS Individual Grant 6G097701.
T.H.K. is supported by a KIAS Individual Grant PG095201 at Korea Institute for Advanced Study.
This work is also supported by the Deutsche Forschungsgemeinschaft under Germany's Excellence Strategy - EXC 2121 Quantum Universe - 390833306.

\appendix

\section{Stability theorems}
\label{app:stab}
We provide theorems for the Q-ball classical stability and their proofs.
Although our argument is based on ref.~\cite{Friedberg:1976me}, we slightly generalize their original argument to apply to our cases.

\subsection{Setup}
Our problem to obtain the Q-ball solution from the Lagrangian \eqref{eq:lagrangian} is rephrased in three different variational problems, called Variational Principle 1, 2, and 3. See \cite{Kusenko:1997ad} for the case of single-field Q-balls.

\paragraph{Variational Principle 1:}
the first one is to minimize the energy of the system~\eqref{FLS Hamiltonian} keeping the charge $Q$.
In this viewpoint, $\omega$ is a functional depending on $Q$ and the fields through eq.~\eqref{eq:Q} and the solutions should satisfy
\begin{align}
 0&= (\delta E[\phi,\chi, \omega[\phi,Q]]) _Q \\[2mm]
&= 
\int d^3 x \left[\frac{\delta E(\phi,\chi, \omega[\phi,Q])}{\delta \chi}\bigg|_{\omega}\delta \chi 
+ \left. \frac{\delta E(\phi,\chi, \omega[\phi,Q])}{\delta \phi} \right|_{\omega}\delta \phi 
+ \frac{\delta E(\phi,\chi, \omega[\phi,Q])}{\delta \omega} \frac{\delta \omega}{\delta \phi} \delta \phi 
\right]\, ,
\end{align}
where the subscripts $Q$ and $\omega$ indicate that they are fixed under the variation or differentiation.
This gives the EOMs~\eqref{eq:QballEOM}.
For the obtained solutions, $E(Q)$ is a function of the input $Q$, and its derivative is given as
\begin{align}
 \frac{d}{d Q} E(Q) & = \frac{ \partial E}{\partial \omega} \frac{\partial \omega}{\partial Q} \\
&= \int d^3 x \, \omega \phi(r)^2  \left(\int \phi(r')^2 d^3 x'\right) ^{-1} = \omega \, , \label{eq:dEdQ}
\end{align}
where we have used that the functional derivatives with respect to the fields vanish since they solve the EOMs.

\paragraph{Variational Principle 2:}
the functional $\mathcal{E}$ defined in eq.~\eqref{modified energy} can be regarded as a Legendre transform of the original energy functional $E$.
In this viewpoint, the fixed input parameter is $\omega$ while $Q$ is a functional of $\omega$ and $\phi$ and is to be varied.
Thus, for the solutions satisfying the Variational Principle 1, $\mathcal{E}$ is a functional of $\chi, \phi$, and $\omega$.
Then one obtains 
\begin{align}
& (\delta \mathcal{E}[\phi,\chi, Q[\phi,\omega]])_\omega \\[2mm]
&=
\left(\delta E[\phi,\chi, Q[\phi,\omega]]\right)_\omega - \omega (\delta Q[\phi,\omega]) _\omega \\[2mm]
&= \int d^3 x \left[
\left.\frac{\delta E(\phi,\chi, Q[\phi,\omega])}{\delta \chi} \right|_{Q} \delta \chi 
+ \left. \frac{\delta E(\phi,\chi, Q[\phi,\omega])}{\delta \phi} \right|_{Q} \delta \phi 
+ \frac{\partial E(\phi,\chi, Q[\phi,\omega])}{\partial Q}  \left.\frac{ \delta Q}{\delta \phi} \right|_\omega\delta \phi
\right] \nonumber \\
&\hspace{1em}
- \omega (\delta Q[\phi,\omega]) _\omega  \\
&=0 \, ,
\end{align}
where the first and second terms vanish because of the Variational Principle 1 and the third term was canceled by the fourth term.
One can easily show that the inverse relation holds, namely, that the solutions satisfying Variational Principle 2 also satisfy Variational Principle 1.
Thus, they are equivalent.
The obtained EOMs are the same as eq.~\eqref{eq:QballEOM}.

\paragraph{Variational Principle 3:}
one may introduce yet another functional $G$:
\begin{align}
 G[\chi,\phi] & \equiv \left. E \right|_{\omega=0} \\
&= \int d^3x \, \left[ \frac{1}{2}(\partial_i \chi)^2 + \frac{1}{2}(\partial_i \phi)^2 + V(\phi,\chi) \right] \equiv \int d^3 x \, \mathcal{G} \, .
\end{align}
$G$ is related to the other functionals as
\begin{equation}
 G = E - \frac{1}{2} \omega Q = \mathcal{E}  + \frac{1}{2} \omega Q \, .
\end{equation}

Minimizing the functional $G$ with a constraint
\begin{align}
 I \equiv \int d^3x \, \phi(r)^2 = I_0 (\mathrm{constant}) \, 
\end{align}
is expressed as
\begin{align}
 (\delta G)_I =0 \Leftrightarrow \delta G [\chi,\phi] - \delta \left(\frac{1}{2}\omega^2 (I-I_0) \right) =0
\end{align}
with $\omega$ being introduced as the Lagrange multiplier.
In this formalism, we treat $\omega$ as an independent variational parameter.

This gives EOMs
\begin{align}
 \frac{\partial G[\chi,\phi]}{\partial \chi} &=0 \label{eq:var3-eom1}\\
 \frac{\partial G[\chi,\phi]}{\partial \phi}  - \frac{1}{2}\omega^2 \frac{\partial I}{\partial\phi}&=0\label{eq:var3-eom2} \\
 \omega I &= \omega I_0 \, ,
\end{align}
which are the same as those given in Variational Principle 1 and 2.

The obtained solutions $\chi,\phi$, and $\omega$ depend on $I$, for which one gets
\begin{align}
 \frac{d }{d I} G(I) = \frac{ \delta G}{\delta \chi} \frac{d \chi}{d I} + \frac{ \delta G}{\delta \phi} \frac{d \phi}{d I}  = \frac{1}{2}\omega^2 \, ,
\end{align}
where we have used \eqref{eq:var3-eom1} and \eqref{eq:var3-eom2}.

\subsection{Hessian}
Under variations with respect to the fields, $\phi+\delta \phi$ and $\chi+\delta \chi$, the second-order variation of the energy with a fixed charge $Q$ is expressed as
\begin{align}
(\delta^2 E)_Q = \frac{1}{2} \int d^3 x \, \psi^t H \psi + \frac{2\omega^3}{Q} \left( \int d^3x \, \psi^t b\right)^2 \label{eq:2nd-var-E}
\end{align}
where
\begin{align} 
 H= - \vec{\partial}\,{}^2 \, 1_{2\times2}+ 
\begin{pmatrix}
\displaystyle \frac{\partial^2 V}{\partial \chi^2}  & \hspace{2em}\displaystyle \frac{\partial^2 V}{\partial \chi \partial \phi} \\[1.5em]
\displaystyle \frac{\partial^2 V}{\partial \chi \partial \phi} & \hspace{2em}\displaystyle \frac{\partial^2 V}{\partial \phi^2}  - 2 \omega^2
\end{pmatrix} \, ,
\end{align}
\begin{equation}
 \psi = 
\begin{pmatrix}
 \delta \chi \\ \delta \phi
\end{pmatrix},
\quad 
b=
\begin{pmatrix}
 0 \\ \phi
\end{pmatrix} \, .
\end{equation}
Similarly, the variations of the other functionals are also expressed as
\begin{align}
 (\delta^2 \mathcal{E})_\omega = \frac{1}{2} \int d^3 x \, \psi^t H \psi  
\end{align} 
and
\begin{equation}
    (\delta^2 G)_I = \frac{1}{2} \int d^3 x \, \psi^t H \psi \qquad \mathrm{with} \quad \int d^3x \, \left[\phi \delta \phi + (\delta \phi) ^2\right]=0~. \label{eq:constraint}
\end{equation} 
We are interested in whether $(\delta^2 E)_Q$ is positive or not under arbitrary perturbations because it corresponds to the criteria of the classical stability of the Q-ball solutions.
Following ref.~\cite{Friedberg:1976me}, we provide several theorems for the stability conditions.

\subsection{Theorem 1}
For a fixed boundary condition, the operator $H$ has at least one negative eigenvalue.\footnote{Note that this is exactly the same as Derrick's theorem~\cite{Derrick:1964ww}.}

\paragraph{Proof.}
It is easiest to prove through Variational Principle 2.
Let the functions $\chi(r)$ and $\phi(r)$ solve the principle $(\delta \mathcal{E})_\omega=0$.
Then consider the scale transformation,
\begin{equation}
 \chi(r) \to \chi(\lambda r) , \quad \phi(r) \to \phi(\lambda r)
\end{equation}
which changes the functional as
\begin{align}
 \mathcal{E} \to \mathcal{E}_\lambda &= \int d^3 x \, \left[ \frac{1}{2}(\partial_i \phi(\lambda r))^2 + \frac{1}{2}(\partial_i \chi(\lambda r))^2 + V(\phi( \lambda r),\chi(\lambda r)) - \frac{1}{2}\omega \phi(\lambda r)^2\right] \\
&= \lambda^{-1} \mathcal{E}_1 + \lambda^{-3} \mathcal{E}_3 \, ,
\end{align}
where
\begin{equation}
 \mathcal{E}_1 \equiv \int d^3 x \, \left[\frac{1}{2}(\partial_i \phi(r))^2 + \frac{1}{2}(\partial_i \chi(r))^2\right], \quad 
 \mathcal{E}_3 \equiv \int d^3 x \, \left[V(\phi( r),\chi( r)) - \frac{1}{2}\omega \phi( r)^2\right] \, .
\end{equation}
Because the solution ensures that the first-derivative of $\mathcal{E}_\lambda$ vanishes, we have
\begin{equation}
0= \left.\frac{d}{d \lambda} \mathcal{E}_\lambda\right|_{\lambda=1}  = - \mathcal{E}_1 -3 \mathcal{E}_3  \Leftrightarrow \mathcal{E}_3 = - \frac{1}{3} \mathcal{E}_1 \, .
\end{equation}
On the other hand, the second derivative is calculated as
\begin{align}
 \left.\frac{d^2}{d^2 \lambda} \mathcal{E}_\lambda\right|_{\lambda=1}  = 2 \mathcal{E}_1 + 12 \mathcal{E}_3 = -2 \mathcal{E}_1 \, ,
\end{align}
which is negative since $\mathcal{E}_1$ is positive.
This means that $H$ has a negative eigenvalue for the eigenfunction
\begin{align}
 \psi = 
\left.
\begin{pmatrix}
\partial_\lambda \chi(\lambda r) \\  \partial_\lambda \phi(\lambda r) 
\end{pmatrix}
\right|_{\lambda=1} \, .
\end{align}

\subsection{Theorem 2}
Assume that there exists some range of $I$, $R_I$, such that the soliton solution of $I\in R_I$ has a lower value of $G$ than that of the plane-wave solution.
Then, the operator $H$ of the soliton solution with the lowest $G$ value (i.e., the branch with the lowest $G$)
has only one negative eigenvalue.

\paragraph{Lemma.}
For an arbitrary $I$, a solution with the lowest $G$ value among the soliton solutions (i.e., the branch with the lowest $G$)
is always the local minimum of the functional $G$ with fixed $I$.

\begin{figure}[tbp]
\centering
\includegraphics[width=0.4\textwidth]{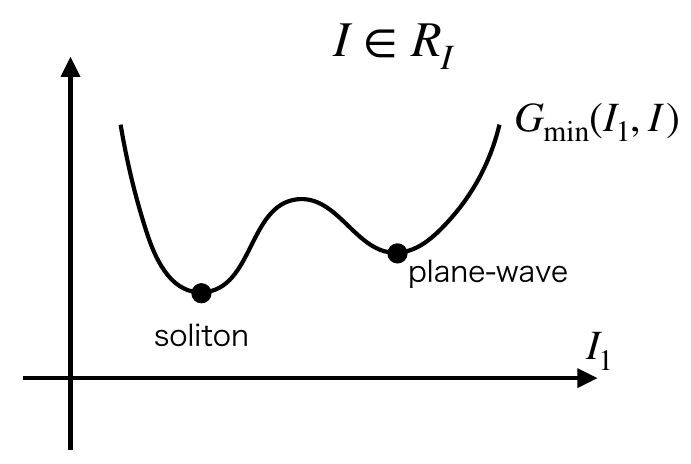}
\includegraphics[width=0.4\textwidth]{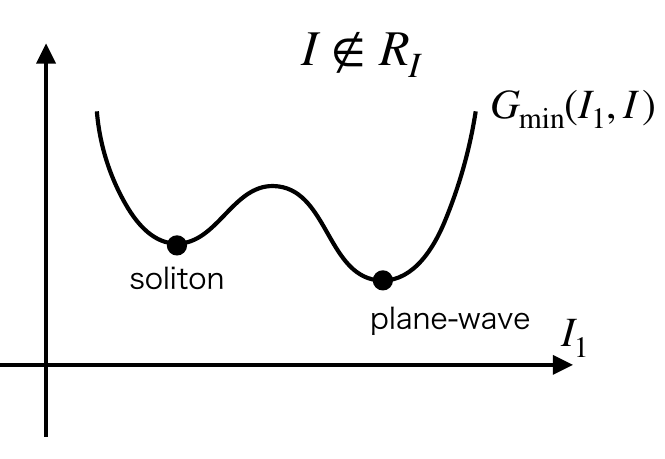} \\[1em]
\includegraphics[width=0.4\textwidth]{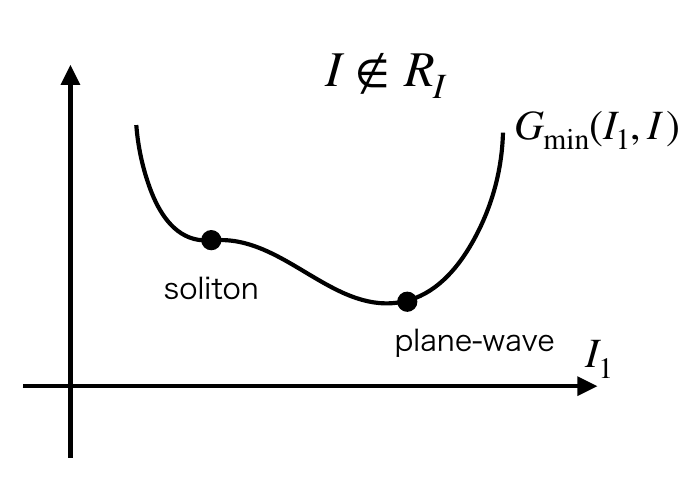}
\caption{Schematic plot of the curve $G_\mathrm{min}(I_1,I)$ vs $I_1$.
All minima in the Variational Principle 3 of $G$ with fixed $I$
must appear as minima in this curve.
Top-left: when $I\in R_I$, the soliton solution has lower $G$ than the plane-wave, being the absolute minimum.
Top-right: when $I\notin R_I$, the soliton solution is no longer the absolute minimum, but remains as local minimum by the lemma.
Bottom: in order for the soliton solution not to be the local minimum, it should merge with another soliton to become degenerated extremum (case (ii)). However, this contradicts the assumption.
}
\label{fig:Gmin}
\end{figure}


\paragraph{Proof of lemma.}
Let $I_1$ be the following functional:
\begin{equation}
 I_1 \equiv \int d^3x\, \phi(r) \, .
\end{equation}
Then consider the minimization problem of $G$ with $I_1$ and $I$ fixed, i.e., $G_{\rm min}(I_1,I)$.

Note that the soliton solutions and the plane-wave solution always have distinct values of $I_1$,
\begin{equation}
 \text{soliton sol} \to I_1 =\mathcal{O}(1), \quad \text{plane-wave sol} \to I_1 = \mathcal{O}(V_3^{1/2}),
\end{equation}
with $V_3$ being the volume of the three-dimensional space.

Let us consider two cases, $I\in R_I$ and $I\notin R_I$.
For $I\in R_I$, the soliton solution has a lower $G$ than that of the plane-wave solution, i.e., the soliton must be the absolute minimum of $G$, which is of course a local minimum as well.
Note that all minima of $G$ with fixed $I$ must appear as minima on the curve $G(I_1,I)$ vs $I_1$.
Therefore, when one draws a plot of the curve $G_{\rm min}(I_1,I)$ vs $I_1$ (for fixed $I$), the soliton and plane-wave solutions are separated minima in which the soliton solution has the lowest $G$. (See the top-left panel in figure~\ref{fig:Gmin}.)
Then, vary $I$ to be $I\notin R_I$.
The soliton solution is no longer the absolute minimum, but remains as a local minimum on the curve $G_{\rm min}(I_1,I)$ vs $I_1$ because otherwise the soliton must merge with either of (i) plane-wave solution or (ii) other soliton solutions, where the case (i) contradicts the fact that they have different values of $I_1$ and the case (ii) implies the lowest branch merging with an upper branch and contradicts the assumption that the soliton solution has the lowest value of $G$ among the solitons.
(See the bottom panel in figure~\ref{fig:Gmin}.)

\paragraph{Proof of theorem.}
From the Lemma, it follows that the soliton solution with the lowest $G$ value is the local minimum of the functional $G$ with fixed $I$.

Let us prove the theorem by contradiction.
Suppose that the corresponding $H$ have two or more negative eigenvalues and denote two of them by $\psi_1$ and $\psi_2$ (with eigenvalues $\lambda_1$ and $\lambda_2$).
By considering a suitable linear combination of them, $\psi_0= c_1 \psi_1 + c_2\psi_2$, we can satisfy the constraint \eqref{eq:constraint}, i.e., $\psi_0$ does not change $I$.
In addition, it is easy to show that the variation in the direction of $\psi_0$ lowers $G$ as
\begin{equation}
 (\delta^2 G)_I = \frac{1}{2}\int d^3x \, \psi_0 ^ t H \psi_0 = \frac{1}{2}\left(c_1^2 \lambda_1 + c_2^2 \lambda_2 \right) < 0 \, .
\end{equation}
Therefore, this contradicts the lemma that the soliton solution is the local minimum of $G$ with fixed $I$.
This proves the theorem.

\subsection{Theorem 3}
The necessary and sufficient conditions for $(\delta^2 E)_Q\geq 0$ under arbitrary perturbations of $\phi$ and $\chi$ are
\begin{equation}
\begin{cases}
\text{(i) $H$ has only one negative eigenvalue} \\
\text{(ii)}~\displaystyle \frac{1}{Q}\frac{d Q}{d \omega} <0 \, .
\end{cases}
\label{thm-stability}
\end{equation}

\paragraph{Proof.}
Let us consider the following function
\begin{equation}
 a \equiv \frac{\partial}{\partial \omega}
\begin{pmatrix}
\chi \\ \phi
\end{pmatrix}
\end{equation}
where $\chi$ and $\phi$ are solutions in the Variational Principle 2 and regarded as functions of $\omega$.
Differentiating the EOMs~\eqref{eq:QballEOM} with respect to $\omega$, one gets 
\begin{equation}
 H a = 2 \omega b \, .\label{eq:cond-a}
\end{equation}
Using the eigenfunctions of $H$, which form an orthonormal complete set $\{\psi_i \}$ with eigenvalues $\lambda_i$, $a$ and $b$ are expressed as
\begin{align}
 a = \sum _i a_i \psi_i  \,,  \quad b = \sum _i b_i \psi_i \, .\label{eq:a-b}
\end{align}
From \eqref{eq:cond-a}, we have
\begin{align}
 a_i \lambda_i = 2 \omega b_i \, ~ \text{(no sum for $i$)}. \label{eq:a_i}
\end{align}
Note that $b_i=0$ for $\lambda_i=0$.

Then consider the following quantity
\begin{align}
 \int d^3x \, a^t H a &=  \sum_i \lambda_i a_i^2  
\end{align}
where we have used eq.~\eqref{eq:a-b}.
Here the l.h.s. is also equal to 
\begin{align}
 \int d^3x \, a^t H a& =  2 \omega  \int d^3x \, a^t b \\
& =  2 \omega  \int d^3x \, \phi \,\partial _\omega \phi \, ,
\end{align}
where we have substituted eq.~\eqref{eq:a_i} and the definition of $a$ into the first and second lines, respectively.
On the other hand, from the definition of $Q$, one has
\begin{align}
 \frac{d}{d \omega} \left(\frac{Q[\phi,\omega]}{\omega} \right)
&= - \left(\frac{Q[\phi,\omega]}{\omega^2} \right)
+ \frac{1}{\omega} \frac{d}{d \omega} \left(\omega \int d^3 x \, \phi^2 \right) \\
&= \frac{d}{d \omega} \int d^3 x \, \phi^2 \\
&= 2 \int d^3 x \, \phi \, \partial_\omega \phi \, ,
\end{align}
which leads to 
\begin{align}
 \omega \frac{d}{d \omega} \left(\frac{Q[\phi,\omega]}{\omega} \right) = \sum_i \lambda_i a_i^2  
\end{align}
or equivalently,
\begin{align}
  \omega \frac{d}{d \omega} \left(\frac{Q[\phi,\omega]}{\omega} \right) = \sum_i{}' \, \frac{4 \omega^2 b_i^2}{\lambda_i}
\end{align}
where $\sum_i{}'$ means the summation over $\lambda_i \neq0$.

Next, consider an arbitrary variation $\psi$,
\begin{equation}
 \psi= \sum_i c_i \psi_i \, .
\end{equation}
From \eqref{eq:2nd-var-E},
$(\delta^2 E)_Q$ is expressed as
\begin{align}
(\delta^2 E)_Q & =
\frac{1}{2} \sum \lambda_i c_i^2 
+  \frac{2\omega^3}{Q} \sum_{i,j}{}' c_i b_i c_j b_j \\
&=  \frac{1}{2}\sum_{i,j}{}' c_i M_{ij} c_j
\end{align}
with
\begin{equation}
 M_{ij} \equiv \lambda_i \delta_{ij} + \frac{4\omega^3}{Q} b_i b_j \, .
\end{equation}
Now the problem is whether the eigenvalues of the matrix $M_{ij}$ are negative or not.
The eigen equation determining the eigenvalue $z$ is given as
\begin{align}
p(z) \equiv \mathrm{det} (z \delta_{ij} - M_{ij}) =0 \, ,
\end{align}
which is equivalent to 
\begin{align}
\mathcal{J}(z) \equiv \frac{p(z)}{\displaystyle \prod_i{}' (z-\lambda_i)} = 0  \, .
\end{align}
Therefore, the stable solution, i.e., $(\delta^2 E)_Q \geq 0$, corresponds to that $\mathcal{J}(z)$ has all positive roots $z$.
Furthermore, noting that one can rewrite $\mathcal{J}(z)$ as 
\begin{align}
\mathcal{J}(z)=   \sum_{i}{}' \frac{4 \omega^3 b_i^2} {Q(z-\lambda_i)} -1 \, ,
\end{align}
we have
\begin{equation}
 \mathcal{J}(0) = - \frac{\omega}{Q} \frac{d Q}{d \omega} \, .
\end{equation}

\begin{figure}[tbp]
\centering
\includegraphics[width=0.55\textwidth]{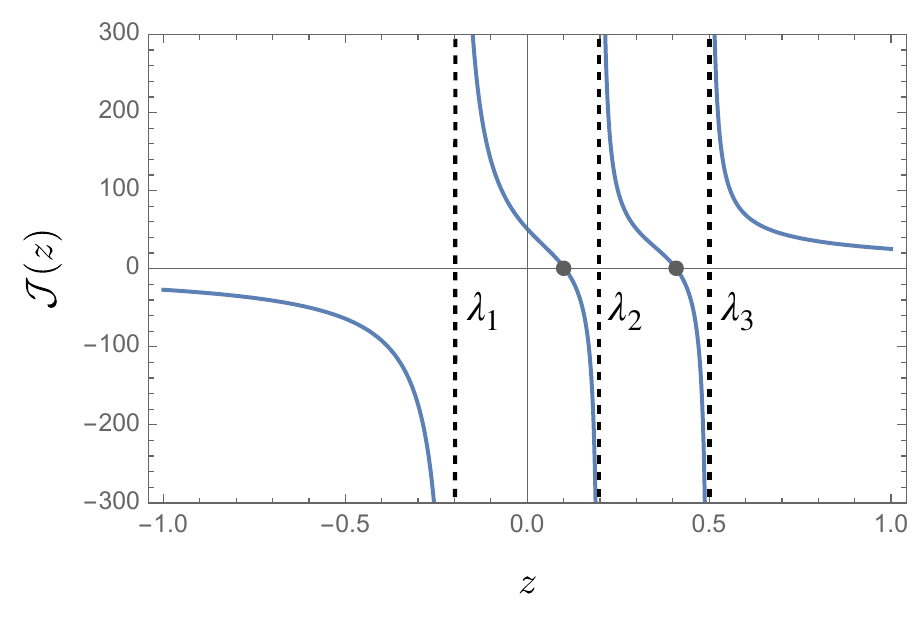}
\caption{Schematic plot of $\mathcal{J}(z)$.
$\mathcal{J}(z) \to -1$ as $z \to  \pm \infty$
and it jumps from $-\infty$ to $\infty$ at $z=\lambda_i$
(vertical dashed lines).
Note that at least one of $\lambda_i$ is negative (Theorem~1).
Roots of $\mathcal{J}(z)=0$ are shown by gray blobs.
}
\label{fig:Jz}
\end{figure}

Then examine the conditions (i) and (ii).
When the conditions are met, only $\lambda_1$ is negative and $\mathcal{J}(0)>0$, leading to figure~\ref{fig:Jz} showing the plot of $\mathcal{J}(z)$, which means that all the roots are positive.
On the other hand, if all roots are positive, the condition (i) is necessary.
(Note that at least one of $\lambda_i$ must be negative, which follows from Theorem~1.)
Obviously the condition (ii) ($\mathcal{J}(0)>0$) is also necessary from figure~\ref{fig:Jz}.
Then Theorem~3 has been proved.

\section{Approximate solutions for thin/thick-wall regimes}

In the thin-wall regime, we can find the limiting charge $Q_{\rm max}$ by directly solving for $\omega_\Lambda$ in eq.~\eqref{eq:dEdomegathinwall},
\begin{equation}
\label{eq:volsolunstable}
    \omega_\Lambda^2 = \frac{\Lambda^2}{4\kappa} {}_4F_3(\tfrac{1}{5},\tfrac{2}{5},\tfrac{3}{5},\tfrac{4}{5};\tfrac{1}{2},\tfrac{3}{4},\tfrac{5}{4};c^{\rm thin})
\end{equation}
for the unstable (local maximum in energy) branch and
\begin{align}
\begin{split}
\label{eq:volsol}
    \omega_\Lambda^2 =& \frac{\Lambda^2}{4\kappa} \left[a^{-1}{}_4F_3(\tfrac{-1}{20},\tfrac{3}{20},\tfrac{7}{20},\tfrac{11}{20};\tfrac{1}{4},\tfrac{1}{2},\tfrac{3}{4};c^{\rm thin}) -\frac{1}{4}{}_4F_3(\tfrac{1}{5},\tfrac{2}{5},\tfrac{3}{5},\tfrac{4}{5};\tfrac{1}{2},\tfrac{3}{4},\tfrac{5}{4};c^{\rm thin})~\right. \\ & \left. - \frac{5}{32}a~{}_4F_3(\tfrac{9}{20},\tfrac{13}{20},\tfrac{17}{20},\tfrac{21}{20};\tfrac{3}{4},\tfrac{5}{4},\tfrac{3}{2};c^{\rm thin}) -\frac{5}{32}a^2~{}_4F_3(\tfrac{7}{10},\tfrac{9}{10},\tfrac{11}{10},\tfrac{13}{10};\tfrac{5}{4},\tfrac{3}{2},\tfrac{7}{4};c^{\rm thin}) \right]
\end{split}
\end{align}
for the stable (local minimum in energy) branch, where
\begin{align}
\begin{split}
\label{eq:cthin}
    &a =  \frac{\sqrt{6Q}\Lambda^2}{4\pi^2 \kappa \sqrt{\lambda}}\left|\frac{\Lambda^2}{4\kappa^2}-v^2\right|^{-1}~,\\
    &c^{\rm thin} = \frac{3125a^4}{256} = \frac{28125 Q^2 \Lambda^8}{16384\pi^8\kappa^4\lambda^2}\left(\frac{\Lambda^2}{4\kappa^2}-v^2\right)^{-4}~.
\end{split}
\end{align}
The hypergeometric functions ${}_4F_3$ in eqs.~\eqref{eq:volsolunstable} and~\eqref{eq:volsol} are $1$ for $c^{\rm thin}=0$ ($Q=0$) and change slightly $\sim 1$ for $c^{\rm thin}=1$ ($Q=Q_{\rm max}$). For $c^{\rm thin}>1$ ($Q>Q_{\rm max}$), they become imaginary, suggesting an instability and reinforcing eq.~\eqref{eq:Qmaxthinwall} as the maximum stable charge. The first term in the stable solution eq.~\eqref{eq:volsol} dominates for small charge, with $\omega_{\Lambda}^2 \sim (\Lambda^2/4\kappa) a^{-1} \xrightarrow{} \infty$ as $Q\xrightarrow{} 0$. The solution steadily decreases with increasing charge and approaches $\omega_\Lambda^2 = 1.237 \Lambda^2/4\kappa$ at $Q=Q_{\rm max}$.

We can also solve directly for $\omega_\Lambda$ in the thick-wall limit. We show this explicitly for moderate values of $1/4<\Lambda/2\kappa v<2\pi$, where we solve for $\omega_\Lambda$ by considering only the $\omega Q$ term in combination with the $a_1$ surface energy term. The two remaining solutions after restricting our parameter space to real positive $\omega_\Lambda$ are
\begin{equation}
\label{eq:wlambdathick}
    \omega_\Lambda^2 = \frac{\Lambda^2}{8\kappa}\left(\sqrt{\Omega}\pm \sqrt{\frac{2}{c\sqrt{\Omega}}-\Omega}~\right)~,
\end{equation}
with
\begin{equation}
    \Omega = \frac{(\Delta + 9c)^{1/3}}{(18)^{1/3} c}+ \frac{4 (2/3)^{1/3}}{(\Delta + 9c)^{1/3}},
\end{equation}
\begin{equation}
    \Delta = \sqrt{3}\sqrt{27c^2-256c^3}~, \label{eq:Delta}
\end{equation}
and
\begin{equation}
    c = \frac{Q^2\Lambda^6}{1024\pi^6 v^4 m_\chi^2 \kappa^3 a_1^2}~.
\end{equation}
From eq.~\eqref{eq:Delta}, we see that the range of $c$ is between $0$ and $27/256$, with this latter value equivalent to the $Q_{\rm max}$ found in eq.~\eqref{eq:qmaxsurf}. 

Of the two solutions, the $-$ solution has $\omega_\Lambda^2 \xrightarrow{} \Lambda^2/4\kappa$ for $c\xrightarrow{} 0$ ($Q\xrightarrow{}0$) and $\omega_\Lambda^2 \xrightarrow{} \Lambda^2/3\kappa$ for $c=27/256$, whereas the $+$ solution approaches infinity for $c=0$ and meets the $-$ solution at $c=27/256$. The $+$ solution is the local minimum in energy and therefore the stable solution.

If $m_\chi/\omega_{\Lambda}\lesssim 1/2\pi$, then the $a_2$ term in the surface energy may play the dominant role in determining $\omega$ via eq.~\eqref{eq:dedwtotal}. Solving directly again for $\omega_\Lambda$, the stable, positive real solution is
\begin{equation}
    \omega_\Lambda^2 = \frac{\Lambda^2}{2\kappa} \left(\sqrt{\frac{1}{3c_2}}\cos\left[\frac{1}{3}\arccos\left(-\frac{3}{2}\sqrt{3c_2}\right)\right]\right)~,
\end{equation}
taking the positive root for $\omega_\Lambda$ with
\begin{equation}
    c_2 = \frac{Q^2 \Lambda^4}{256\pi^4v^4\kappa^2 a_2^2}~.
\end{equation}
For small values of $Q$ (where the $a_2$ surface term would more often dominate the $a_1$ term), the leading terms in the expansion are $\omega_\Lambda^2 \xrightarrow{}(\Lambda^2/4\kappa)(\sqrt{1/c_2}-1/2)$.

\section{Minimum charge estimate} \label{app:Qmin}

We analytically estimate the minimum stable charge $Q_{\rm min}$. Since stable Q-ball solutions satisfy eq.~\eqref{thm-stability}, $Q_{\rm min}$ occurs at the largest $\omega$ for the stable branch, near $\omega = m_\Phi$ as can be seen in figure~\ref{fig:Q-E-omega}. Although the analytic solution is not strictly valid for $\omega\xrightarrow{} m_\Phi$, we estimate $Q_{\rm min}$ by setting $\omega=m_\Phi$ ($\omega_\Lambda =\sqrt{m_\Phi^2 + \Lambda^2/4\kappa}$) in eq.~\eqref{eq:dedwtotal} and solving for charge:
\begin{equation}
\label{eq:qminapprox}
    Q_{\rm min} \approx \frac{\lambda \pi^4 m_\Phi}{6(m_\Phi^2+\Lambda^2/4\kappa)^{5/2}}\left(\frac{\Lambda^2}{4\kappa^2}-v^2\right)^2 + \frac{4\pi^3v^2m_\chi m_\Phi a_1}{(m_\Phi^2+\Lambda^2/4\kappa)^2}+\frac{4\pi^2v^2m_\Phi a_2}{(m_\Phi^2+\Lambda^2/4\kappa)^{3/2}}~.
\end{equation}
Eq.~\eqref{eq:qminapprox} generally reproduces the numerical behavior found in figure~\ref{fig:LambdaQplane}, albeit consistently overestimating the minimum charge.



\bibliography{qballs}
\addcontentsline{toc}{section}{Bibliography}
\bibliographystyle{JHEP}
\end{document}